# Improved Approximation of Linear Threshold Functions[*]

Ilias Diakonikolas[†]
Department of Computer Science
Columbia University
New York, NY
ilias@cs.columbia.edu

Rocco A. Servedio[‡]
Department of Computer Science
Columbia University
New York, NY
rocco@cs.columbia.edu

October 29, 2018

**Abstract**

We prove two main results on how arbitrary linear threshold functions $f(x) = \text{sign}(w \cdot x - \theta)$ over the $n$-dimensional Boolean hypercube can be approximated by simple threshold functions.

Our first result shows that every $n$-variable threshold function $f$ is $\epsilon$-close to a threshold function depending only on $\text{Inf}(f)^2 \cdot \text{poly}(1/\epsilon)$ many variables, where $\text{Inf}(f)$ denotes the total influence or average sensitivity of $f$. This is an exponential sharpening of Friedgut's well-known theorem [Fri98], which states that every Boolean function $f$ is $\epsilon$-close to a function depending only on $2^{O(\text{Inf}(f)/\epsilon)}$ many variables, for the case of threshold functions. We complement this upper bound by showing that $\Omega(\text{Inf}(f)^2 + 1/\epsilon^2)$ many variables are required for $\epsilon$-approximating threshold functions.

Our second result is a proof that every $n$-variable threshold function is $\epsilon$-close to a threshold function with integer weights at most $\text{poly}(n) \cdot 2^{\tilde{O}(1/\epsilon^{2/3})}$. This is a significant improvement, in the dependence on the error parameter $\epsilon$, on an earlier result of [Ser07] which gave a $\text{poly}(n) \cdot 2^{\tilde{O}(1/\epsilon^2)}$ bound. Our improvement is obtained via a new proof technique that uses strong anti-concentration bounds from probability theory. The new technique also gives a simple and modular proof of the original [Ser07] result, and extends to give low-weight approximators for threshold functions under a range of probability distributions beyond just the uniform distribution.

---

[*]A preliminary version of this work appeared in the *Proceedings of the 24th Annual IEEE Conference on Computational Complexity (CCC)* [DS09].

[†]Research supported by NSF grants CCF-0728736, CCF-0525260, and by an Alexander S. Onassis Foundation Fellowship.

[‡]Supported in part by NSF grants CCF-0347282, CCF-0523664 and CNS-0716245, and by DARPA award HR0011-08-1-0069.

# 1 Introduction

Linear threshold functions (henceforth simply called *threshold functions*) are functions $f : \{-1, 1\}^n \to \{-1, 1\}$ of the form $f(x) = \text{sign}(w \cdot x - \theta)$ where the weights $w_1, \ldots, w_n$ and the threshold $\theta$ may be arbitrary real values. Threshold functions are a fundamental type of Boolean function and have played an important role in computer science for decades, see e.g. [Der65, Mur71, SRK95]. Recent years have witnessed a flurry of research activity on threshold functions from many perspectives of theoretical computer science, including hardness of learning [FGKP06, KS08], efficient learning algorithms in various models [Kal07, OS08, KKMS08], property testing [MORS09, GS07], communication complexity and circuit complexity [She07], monotone computation [BW06], derandomization [RS09, DGJ$^+$09], and more.

Despite their seeming simplicity threshold functions can have surprisingly rich structure, and basic questions about them can be unexpectedly challenging to answer. As one example, a moment's thought shows that every threshold function $f$ can be realized with integer weights $w_1, \ldots, w_n$: how large do those integer weights need to be? A fairly straightforward argument gives a bound of $2^{O(n \log n)}$, but while this upper bound was known at least since 1961 [MTT61] and rediscovered several times (e.g. [Hon87, Rag88]), more than thirty years elapsed before a matching lower bound of $2^{\Omega(n \log n)}$ was finally obtained via a fairly sophisticated construction and proof [Hås94, AV97].

This paper is about approximating arbitrary threshold functions using "simple" threshold functions, meaning ones that depend on few variables or have small integer weights. We use a natural notion of approximation with respect to the uniform distribution: throughout the paper "$h$ is an $\epsilon$-approximator for $f$" means that $\mathbf{Pr}[h(x) \neq f(x)] \leq \epsilon$. (All probabilities and expectations over $x \in \{-1, 1\}^n$ are taken with respect to the uniform distribution, unless otherwise specified. In Section 4 we shall consider more general notions of approximation with respect to other distributions as well.) We prove two main results about approximating threshold functions, which we motivate and describe below.

## 1.1 First main result: optimally approximating threshold functions by juntas.

The *influence* of coordinate $i$ on $f : \{-1, 1\}^n \to \{-1, 1\}$ is $\text{Inf}_i(f) \stackrel{\text{def}}{=} \mathbf{Pr}[f(x) \neq f(x^{\oplus i})]$, where $x^{\oplus i}$ denotes $x$ with the $i$-th bit flipped. The *total influence* of $f$, written $\text{Inf}(f)$, is $\sum_i \text{Inf}_i(f)$; it is a normalized measure of the fraction of edges in the hypercube that are rendered bichromatic by $f$, and is equal to the "average sensitivity" of $f$. It is well known (see [FK96] or [BT96] for an explicit proof) that every threshold function has $\text{Inf}(f) \leq \sqrt{n}$, and that the majority function on $n$ variables achieves $\text{Inf}(f) = \Theta(\sqrt{n})$ – and in fact maximizes $\text{Inf}(f)$ over all threshold (or even all unate) functions.

In [Fri98], Friedgut proved the following:

**Theorem.** *[Fri98] Every Boolean function $f$ is $\epsilon$-approximated by a $2^{O(\text{Inf}(f)/\epsilon)}$-junta, i.e. a function depending only on $2^{O(\text{Inf}(f)/\epsilon)}$ of the $n$ input variables.*

Friedgut's theorem is an important structural result about boolean functions and has been usefully applied in several areas of theoretical computer science, including hardness of approx-imation [DS05, CKK+06, KR08], metric embeddings [KR06], and learning theory [OS07]. In Section 2.5 we discuss the role of this theorem in a sequence of results on the Fourier representation of Boolean functions.

Friedgut showed that his bound is best possible for general Boolean functions, by giving an explicit family of functions which require $2^{\Omega(\mathrm{Inf}(f)/\epsilon)}$-juntas for any $\epsilon$-approximation. A bound of the form $2^{O(\mathrm{Inf}(f)/\epsilon)}$ is of course nontrivial only if $\mathrm{Inf}(f) \ll \log n$, which is rather small; thus, it is natural to ask whether various restricted classes of functions, such as threshold functions, might admit stronger bounds.

Our first main result is an exponentially stronger version of Friedgut's theorem for threshold functions:

**Theorem 1** (First Main Theorem). *Every threshold function $f$ is $\epsilon$-approximated by an $\mathrm{Inf}(f)^2 \cdot \mathrm{poly}(1/\epsilon)$-junta (which is itself a threshold function).*

This bound is essentially optimal; easy examples show that $\Omega(\mathrm{Inf}(f)^2 + 1/\epsilon^2)$ many variables may be required for $\epsilon$-approximation (see Section 2.5). We conjecture that Theorem 1 extends to degree-$d$ polynomial threshold functions with an exponential dependence on $d$ in the bound, and also conjecture a different extension of Theorem 1 that is inspired by a theorem of Bourgain; see Section 2.5.

**Techniques.** The proof of Friedgut's theorem makes essential use of the Bonami-Gross-Beckner hypercontractive inequality [Bon70, Gro75, Bec75]. Our proof of Theorem 1 takes a completely different route and does not use hypercontractivity; instead, the main ingredients are recent Fourier results on threshold functions from [OS08] and a probabilistic construction which is reminiscent of Bruck and Smolensky's randomized construction of polynomial threshold functions [BS92].

In more detail, a key notion in our proof is that of a *regular* threshold function; roughly speaking, this is a threshold function where each of the weights $w_i$ is "small" relative to the 2-norm of the weight vector. Given a regular threshold function $g = \mathrm{sign}(w \cdot x - \theta)$, we use the weights $w_i$ to define a probability distribution over approximators to $g$ (this is done similarly to [BS92]). We show (Lemmas 8 and 9) that a randomly drawn approximator from this distribution has high expected accuracy and does not depend on too many variables (the upper bound is given in terms of the weights $w_i$ and the regularity parameter).

An obvious problem in using this construction to approximate arbitrary threshold functions is that not every threshold function is regular. To get around this, we use a recent result from [OS08] which shows that every threshold function $f$ can be well approximated by a threshold function $f'$ which has two crucial properties: $f'$ is *almost* regular (in the sense that it only has a few "large" weights), and its "small" weights are (appropriately scaled versions of) the influences of the corresponding variables in $f$. For each restriction $\rho$ that fixes the large-weight variables of $f'$, then, we may use $f'|_\rho$ as the regular threshold function $g$ of the previous paragraph, and we obtain a distribution over approximators to $f'|_\rho$ where the number of relevant variables for each such approximator is at most $\mathrm{Inf}(f)^2 \cdot \mathrm{poly}(1/\epsilon)$. From this, using the probabilistic method, we are able to argue that there is a *single* high-accuracy approximator for $f$ that depends on at most $\mathrm{Inf}(f)^2 \cdot \mathrm{poly}(1/\epsilon)$ variables, as required.

## 1.2 Second main result: approximating threshold functions to higher accuracy.

The second main result of this paper is about approximating an arbitrary $n$-variable threshold function $f$ using a threshold function $g$ with *small integer weights*. Goldberg [Gol06] and Servedio [Ser07] have observed that, because of the $2^{\Omega(n \log n)}$ lower bound [Hås94] on integer weights to exactly represent arbitrary threshold functions, it is not possible in general to construct an $\epsilon$-approximator $g$ with integer weights $\text{poly}(n, 1/\epsilon)$. Servedio [Ser07] gave the first positive result, by showing that for every threshold function $f$ there is an $\epsilon$-approximating threshold function $g$ in which each weight is an integer of magnitude at most $\text{poly}(n) \cdot 2^{\tilde{O}(1/\epsilon^2)}$. This result and the ingredients in its proof have since played an important role in subsequent work on threshold functions, e.g. [OS08, MORS09, DGJ$^+$09].

Given the usefulness of [Ser07] and the poor dependence on $\epsilon$ in its bound, it is natural to seek a stronger quantitative bound with a better dependence on $\epsilon$; in fact, this was posed as a main open question in [Ser07]. Our second main result makes progress in this direction:

**Theorem 2** (Second Main Theorem). *Every $n$-variable threshold function $f$ is $\epsilon$-approximated by a threshold function $g = \text{sign}(w \cdot x - \theta)$ with $w_1, \ldots, w_n$ all integers of magnitude $n^{3/2} \cdot 2^{\tilde{O}(1/\epsilon^{2/3})}$.*

Another question posed in [Ser07] asked about small integer-weight approximators with respect to other probability distributions beyond just the uniform distribution. As described below, Theorem 2 can be generalized to hold under a range of non-uniform distributions.

Theorem 2 is proved using a new approach which we believe may lead to better bounds for a range of problems considered in [OS08, MORS09, DGJ$^+$09] which use the approach from [Ser07]. Roughly speaking, the proof in [Ser07] and the applications in [OS08, MORS09, DGJ$^+$09] all rely on the fact that for suitable weight vectors $w$, the random variable $w \cdot x$ (with $x$ uniform over $\{-1, 1\}^n$) can be approximated by a Gaussian. Such approximation provides a great deal of information about $w \cdot x$, but the drawback is that the Gaussian is only a fairly coarse approximator of $w \cdot x$ even for a weight vector as well-behaved as $w = (1, \ldots, 1)$, and this inevitably seems to lead to bounds that are exponential in $1/\epsilon^2$ (as in [Ser07, OS08, DGJ$^+$09]). We now briefly describe how our new approach that yields Theorem 2 gets around this barrier.

**Techniques.** The main conceptual difference between our new approach and the approach in [Ser07] is this. The proof in [Ser07] starts with an *arbitrary* vector of weights that represent some threshold function; intuitively this could be problematic because these weights may provide an inconvenient representation to work with for the underlying function. In contrast, we focus on the *function itself*, and prove that every threshold function has a "nice" weight vector that represents it. This allows us to exploit strong anti-concentration bounds [Hal77] that apply only under certain assumptions on the weights; we elaborate below.

The notion of *anti-concentration* is an important ingredient in our approach: a random variable has good anti-concentration if it does not assign too much mass to any small interval of the real line. The study of anti-concentration has a rich history in probability theory, see e.g. [DL36, Kol60, Ess68, Rog73, RV08]. Anti-concentration inequalities for discrete random variables of the form $w \cdot x$ are known to be significantly more delicate than concentration inequalities (i.e. "tail bounds"): while concentration typically depends on the 2-norm of $w$, anti-concentration depends on the *additive structure* of the coefficients in a subtle way.[1]

We remark that [Ser07] also (implicitly) uses anti-concentration bounds, in particular ones based on Gaussian approximation (that follow from the Berry-Esséen Theorem; see Theorem 4). In hindsight it can be seen that no stronger anti-concentration bounds can be used in the arguments of [Ser07] because that proof considers *all possible* representations of the form $\text{sign}(w \cdot x - \theta)$, where $w$ ranges over all of $\mathbb{R}^n$. As an example, consider the majority function. For the standard representation as $\text{sign}(\sum_i x_i)$, the anti-concentration bound given by the Berry-Esséen Theorem is the best possible, since an arbitrarily small interval that contains the origin has probability mass $\Omega(1/\sqrt{n})$. On the other hand, it is possible to come up with alternate representations $\text{sign}(w \cdot x)$ for the majority function that have better anti-concentration; this is essentially what our proof does. We prove a structural theorem which states that every threshold function has a representation in which "many" weights are "well-separated;" under this condition on the weights, we obtain strong anti-concentration using a result of of Halász [Hal77]. Finally, we show that strong anti-concentration yields low-weight integer approximation to get our final desired result.

**Discussion:** Our general approach is both modular and robust. It yields a simple and modular proof of the $\text{poly}(n) \cdot 2^{\tilde{O}(1/\epsilon^2)}$ upper bound from [Ser07] which was proved there via a rather elaborate case analysis. More importantly, the new $\text{poly}(n) \cdot 2^{\tilde{O}(1/\epsilon^{2/3})}$ bound and its proof generalize easily to a wide range of distributions. These include constant-biased product distributions and, using the recent result of [DGJ+09], all $K$-wise independent distributions for sufficiently large $K$ ($K = \tilde{O}(1/\epsilon^2)$ suffices for $\epsilon$-approximation).

**Organization.** We prove Theorem 1 in Section 2 and Theorem 2 in Section 3. Section 4 contains the extension of Theorem 2 to certain nonuniform distributions.

## 2 Theorem 1: Optimally approximating threshold functions by juntas

This section is structured as follows: after giving some mathematical preliminaries, in Section 2.2 we describe a randomized construction of approximators for regular threshold functions. In Section 2.3 we recall the result from [OS08] that lets us approximate any threshold function by a threshold function that is "almost" regular. In Section 2.4 we put these pieces together to prove Theorem 1. We give some discussion and conjectures in Sections 2.5.

### 2.1 Preliminaries.

#### 2.1.1 Basic Probabilistic Inequalities.

We first recall the following standard additive Hoeffding bound:

---

[1] Roughly speaking, if one forbids more and more additive structure in the $w_i$'s, then one gets better and better anti-concentration; see e.g. [Vu08, TV08] and Chapter 7 of [TV06].

**Theorem 3.** *Let $X_1, \ldots, X_n$ be independent random variables such that for each $j \in [n]$, $X_j$ is supported on $[a_j, b_j]$ for some $a_j, b_j \in \mathbb{R}$, $a_j \leq b_j$. Let $X = \sum_{j=1}^n X_j$. Then, for any $t > 0$,*

$$\mathbf{Pr}\left[|X - \mathbf{E}[X]| \geq t\right] \leq 2 \exp\left(\frac{-2t^2}{\sum_{j=1}^n (b_j - a_j)^2}\right).$$

The Berry-Esséen theorem is a version of the Central Limit Theorem with explicit error bounds:

**Theorem 4.** *(Berry-Esséen) Let $X_1, \ldots, X_n$ be independent random variables satisfying $\mathbf{E}[X_i] = 0$ for all $i \in [n]$, $\sqrt{\sum_i \mathbf{E}[X_i^2]} = \sigma$, and $\sum_i \mathbf{E}[|X_i|^3] = \rho_3$. Let $S = (X_1 + \cdots + X_n)/\sigma$ and let $F$ denote the cumulative distribution function (cdf) of $S$. Then*

$$\sup_x |F(x) - \Phi(x)| \leq C\rho_3/\sigma^3,$$

*where $\Phi$ is the cdf of a standard Gaussian random variable, and $C$ is a universal constant. [Shi86] has shown that one can take $C = .7915$.*

**Corollary 5.** *Let $x_1, \ldots, x_n$ denote independent uniformly random $\pm 1$ signs and let $w_1, \ldots, w_n \in \mathbb{R}$. Write $\sigma = \sqrt{\sum_i w_i^2}$, and assume $|w_i|/\sigma \leq \tau$ for all $i \in [n]$. Then for any interval $[a,b] \subseteq \mathbb{R}$,*

$$\left|\mathbf{Pr}[a \leq w_1 x_1 + \cdots + w_n x_n \leq b] - \Phi([\tfrac{a}{\sigma}, \tfrac{b}{\sigma}])\right| \leq 2\tau,$$

*where $\Phi([c,d]) \stackrel{\text{def}}{=} \Phi(d) - \Phi(c)$. In particular,*

$$\mathbf{Pr}[a \leq w_1 x_1 + \cdots + w_n x_n \leq b] \leq \frac{|b-a|}{\sigma} + 2\tau.$$

### 2.1.2 Fourier Analysis over $\{-1,1\}^n$.

We consider functions $f : \{-1,1\}^n \to \mathbb{R}$ (though we often focus on Boolean-valued functions which map to $\{-1,1\}$), and we think of the inputs $x$ to $f$ as being distributed according to the uniform probability distribution. The set of such functions forms a $2^n$-dimensional inner product space with inner product given by $\langle f, g \rangle = \mathbf{E}[f(x)g(x)]$. The set of functions $(\chi_S)_{S \subseteq [n]}$ defined by $\chi_S(x) = \prod_{i \in S} x_i$ forms a complete orthonormal basis for this space. We will often simply write $x_S$ for $\prod_{i \in S} x_i$.

Given a function $f : \{-1,1\}^n \to \mathbb{R}$ we define its *Fourier coefficients* by $\widehat{f}(S) \stackrel{\text{def}}{=} \mathbf{E}[f(x)x_S]$, and we have that $f(x) = \sum_S \widehat{f}(S) x_S$. We refer to the maximum $|S|$ over all nonzero $\widehat{f}(S)$ as the *Fourier degree* of $f$. When $|S| = 1$ we usually abuse notation and write $\widehat{f}(i)$ instead of $\widehat{f}(\{i\})$.

As an easy consequence of orthonormality we have *Plancherel's identity* $\langle f, g \rangle = \sum_S \widehat{f}(S) \widehat{g}(S)$, which has as a special case *Parseval's identity*, $\mathbf{E}[f(x)^2] = \sum_S \widehat{f}(S)^2$. From this it follows that for every $f : \{-1,1\}^n \to \{-1,1\}$ we have $\sum_S \widehat{f}(S)^2 = 1$.

We recall the well-known fact (see e.g. [KKL88]) that the total influence $\text{Inf}(f)$ of any Boolean function equals $\sum_S \widehat{f}(S)^2 |S|$. Moreover, for every threshold function $f$ (in fact for every unate function), we have that $\text{Inf}_i(f) = |\widehat{f}(i)|$.

### 2.1.3 Other Technical Preliminaries.

A function $f : \{-1,1\}^n \to \{-1,1\}$ is said to be a "junta on $\mathcal{J} \subseteq [n]$" if $f$ only depends on the coordinates in $\mathcal{J}$. As stated earlier, we say that $f$ is a $J$-junta, $0 \leq J \leq n$, if it is a junta on some set of cardinality at most $J$. For a vector $u \in \mathbb{R}^m$ we write $\|u\|_1$ to denote the $L_1$ norm of $u$, i.e. $\|u\|_1 = \sum_{i=1}^m |u_i|$. We write "$X \leftarrow \mathcal{D}$" to indicate that random variable $X$ is distributed according to distribution $\mathcal{D}$.

Finally, we give a precise definition of the notion of a "regular" threshold function:

**Definition 6.** *Let $f(x) = \text{sign}(w_0 + \sum_{i=1}^n w_i x_i)$ be a threshold function where $\sum_{i=1}^n w_i^2 = 1$. We say that $f$ is $\tau$-regular if $|w_i| \leq \tau$ for all $i \in [n]$.*[2]

## 2.2 Randomly constructing approximators to regular threshold functions.

Fix $h_\theta(x) = \text{sign}(\theta + \sum_{i=1}^m u_i x_i)$ to be a $\tau$-regular threshold function, so $\sum_{i=1}^m u_i^2 = 1$ and $|u_i| \leq \tau$ for all $i \in [m]$. Our notation emphasizes the threshold parameter $\theta$ since it will play an important role later.

We begin by defining a distribution $\mathcal{D}$ over linear forms $L(x) = \sum_{i=1}^m c_i x_i$. The distribution $\mathcal{D}$ is defined using the weights $u_i$ similarly to how Bruck and Smolensky [BS92] define a distribution over polynomials using the Fourier coefficients of a Boolean function. A draw of $L(x)$ from $\mathcal{D}$ is obtained as follows: $L(x)$ is first initialized to 0. Then the following is independently repeated $N \stackrel{\text{def}}{=} \Theta(\|u\|_1^2 \cdot \frac{1}{\tau^2} \cdot \ln(1/\tau))$ times: an index $i \in [m]$ is selected with probability $\frac{|u_i|}{\|u\|_1}$, and $\text{sign}(u_i) x_i$ is added to $L(x)$.

Fix any $z \in \{-1,1\}^m$. For $L \leftarrow \mathcal{D}$, we may view $L(z)$ as a sum of $N$ i.i.d. $\pm 1$-valued random variables $Z_1(z), \ldots, Z_N(z)$, where the expectation of each $Z_j(z)$ is $\sum_{i=1}^m \frac{|u_i|}{\|u\|_1} \text{sign}(u_i) z_i = \frac{1}{\|u\|_1} u \cdot z$. We thus have:

$$\mathbf{E}_{L \leftarrow \mathcal{D}}[L(z)] = \sum_{j=1}^N \mathbf{E}[Z_j(z)] = \frac{N}{\|u\|_1}(u \cdot z). \quad (1)$$

With $\mathcal{D}$ in hand we define a distribution $\mathcal{D}'$ over threshold functions $g_\theta$ in the following natural way: to draw a function $g_\theta \leftarrow \mathcal{D}'$ we draw $L \leftarrow \mathcal{D}$ and set

$$g_\theta(x) = \text{sign}(\theta + \frac{\|u\|_1}{N} L(x)). \quad (2)$$

We would like to show that for $g_\theta \leftarrow \mathcal{D}'$, the probability that $g_\theta(z)$ disagrees with $h_\theta(z)$ is "small," i.e. at most $O(\tau)$. But such a bound cannot hold for every $z \in \{-1,1\}^m$, for if the value of $\theta + u \cdot z$ is arbitrarily close to 0 then the expected value of the argument to sign in (2) may be arbitrarily close to 0. For $z$ such that $\theta + u \cdot z$ is not too close to 0, though, it is possible to argue that $g_\theta(z)$ is incorrect only with small probability (over the draw of $g_\theta \leftarrow \mathcal{D}'$). Moreover, the regularity of $h_\theta$ lets us argue that only a small fraction of inputs $z$ have $\theta + u \cdot z$ close to 0, so we can conclude that the expected error of $g_\theta$ is low. We now provide the details.

---

[2] Strictly speaking, $\tau$-regularity is a property of a particular representation $\text{sign}(w_0 + \sum_{i=1}^n w_i x_i)$ and not of a threshold function $f$, which could have different representations some of which are $\tau$-regular and some of which are not. The particular representation we are concerned with will always be clear from context. A similar remark holds for Definition 7.

We will use the following notion of the "margin" of an input relative to a threshold function:

**Definition 7.** *Let $f(x) = \text{sign}(w_0 + \sum_{i=1}^n w_i x_i)$ be a threshold function where the weights are scaled so that $\sum_{i=1}^n w_i^2 = 1$. Given a particular input $z \in \{-1,1\}^n$ we define $\text{marg}(f, z) \stackrel{\text{def}}{=} |w_0 + \sum_{i=1}^n w_i z_i|$.*

Let $\text{MARG}_{\theta, \tau} \stackrel{\text{def}}{=} \{z \in \{-1,1\}^m : \text{marg}(h_\theta, z) \geq \tau\}$ denote the set of points in $\{-1,1\}^m$ with margin at least $\tau$ under $h_\theta$. We now show that a random $g_\theta \leftarrow \mathcal{D}'$ has high expected accuracy on each point $z \in \text{MARG}_{\theta, \tau}$:

**Lemma 8.** *For each $z \in \text{MARG}_{\theta, \tau}$ we have $\mathbf{Pr}_{g_\theta \leftarrow \mathcal{D}'}[h_\theta(z) \neq g_\theta(z)] \leq \tau$. Moreover, each $g_\theta \leftarrow \mathcal{D}'$ is an $N$-junta.*

*Proof.* The latter claim is immediate so it suffices to prove the former. Fix any $z \in \text{MARG}_{\theta, \tau}$, so $|\theta + u \cdot z| \geq \tau$. We need to bound from above the probability of the "bad" event (over the random choice of $g_\theta \leftarrow \mathcal{D}'$) that $h_\theta(z) \neq g_\theta(z)$; we refer to this bad event as $B$.

The key claim is that if $B$ occurs then it must be the case that $|L(z) - \mathbf{E}_{L \leftarrow \mathcal{D}}[L(z)]| \geq \frac{N\tau}{\|u\|_1}$. For suppose that $h_\theta(z) = 1$ and $g_\theta(z) = -1$ (the other case is handled similarly). By definition, we have that $\theta + u \cdot z \geq 0$ and $\theta + (\|u\|_1/N)L(z) < 0$. Since $z$ belongs to $\text{MARG}_{\theta, \tau}$, the first inequality gives that $\theta + u \cdot z \geq \tau$, which implies, via (1), that $\mathbf{E}[L(z)] \geq (N/\|u\|_1)(\tau - \theta)$. The second inequality is equivalent to $L(z) < -\theta N/\|u\|_1$, and consequently we have $\mathbf{E}[L(z)] - L(z) \geq N\tau/\|u\|_1$.

We thus have that $\mathbf{Pr}_{g_\theta \leftarrow \mathcal{D}'}[h_\theta(z) \neq g_\theta(z)] \leq \mathbf{Pr}_{L \leftarrow \mathcal{D}}[|L(z) - \mathbf{E}[L(z)]| \geq \frac{N\tau}{\|u\|_1}]$. Now we again view $L(z)$ as the sum of $N$ i.i.d. $\{-1,1\}$ random variables. The Hoeffding bound (Theorem 3) yields

$$\mathbf{Pr}_{L \leftarrow \mathcal{D}}[|L(z) - \mathbf{E}[L(z)]| \geq \frac{N\tau}{\|u\|_1}] \leq 2\exp\left(-2\frac{(N\tau/\|u\|_1)^2}{4N}\right) \leq \tau,$$

where the second inequality follows by our choice of $N$. This completes the proof of the lemma. □

We next note that by the regularity of $h_\theta$, most points in $\{-1,1\}^m$ have a large margin (and hence are covered by Lemma 8):

**Lemma 9.** $\mathbf{Pr}_{x \in \{-1,1\}^m}[x \notin \text{MARG}_{\theta, \tau}] \leq 4\tau$.

*Proof.* The proof is a consequence of regularity via the Berry-Esséen theorem (see Section 2.1.1); it follows directly by applying (the last statement of) Corollary 5 noting that $\sum_{i=1}^m u_i^2 = 1$. □

Combining Lemmas 8 and 9, we get the main result of this subsection:

**Lemma 10.** $\mathbf{E}_{g_\theta \leftarrow \mathcal{D}'}[\mathbf{Pr}_{x \in \{-1,1\}^m}[g_\theta(x) \neq h_\theta(x)]] \leq 5\tau$.

## 2.3 Approximating threshold functions using their influences as (almost all of) the weights.

Our next tool is the following theorem on approximating threshold functions. Roughly, it says that every threshold function $f$ can be well approximated by a threshold function $f'$ where all but the $\text{poly}(1/\epsilon)$ largest weights of $f'$ have a special structure: up to sign, they are the values $\text{Inf}_i(f)$. (Recall that for a threshold function $f$ we have $|\widehat{f}(i)| = \text{Inf}_i(f)$; see Section 2.1.2.)

**Theorem 11.** *[Theorem 17 of [OS08]] There is a fixed polynomial $\kappa(\epsilon) = \Theta(\epsilon^{144})$[3] such that the following holds: Let $f(x) = \text{sign}\left(w_0 + \sum_{i \in H} w_i x_i + \sum_{i \in T} w_i x_i\right)$ be a threshold function over head indices $H$ and tail indices $T$, where $H \stackrel{\text{def}}{=} \{i : |\widehat{f}(i)| \geq \kappa(\epsilon)^2\}$ and $T$ satisfies $\sum_{i \in T} w_i^2 = 1$. Then either:*

*(i) $f$ is $O(\epsilon)$-close to a junta over $H$; or,*

*(ii) $f$ is $O(\epsilon)$-close to the threshold function $f'(x) = \text{sign}\left(w_0 + \sum_{i \in H} w_i x_i + \sum_{i \in T} \frac{\widehat{f}(i)}{\sigma_T} x_i\right)$,*

*where $\sigma_T$ denotes $\sqrt{\sum_{i \in T} \widehat{f}(i)^2}$. Moreover, in this case we have $\sigma_T = \Omega(\epsilon^2)$.[4]*

Note that $\sum_{i \in T} (\widehat{f}(i)/\sigma_T)^2 = 1$, and since $\sigma_T = \Omega(\epsilon^2)$, for each $i \in T$ we have

$$|\widehat{f}(i)/\sigma_T| < \kappa(\epsilon)^2/\sigma_T \leq O(\epsilon^{288})/\Omega(\epsilon^2) = O(\epsilon^{286}). \tag{3}$$

This means that for any restriction $\rho$ fixing the variables in $H$, the function $f'|_\rho$ is $\text{poly}(\epsilon)$-regular; this is important since it will allow us to apply the results of Section 2.2 to these restrictions.

## 2.4 Proof of Theorem 1.

Now we are ready to prove Theorem 1. We first show that every threshold function $f$ is $O(\epsilon)$-approximated by a $(1 + \text{Inf}(f)^2) \cdot \text{poly}(1/\epsilon)$-junta threshold function, and then argue that this yields Theorem 1. For brevity, in the rest of this subsection we write $\mathbb{I}$ for $\text{Inf}(f)$.

Let $0 < \epsilon < \frac{1}{2}$ be given and let $f$ be any $n$-variable threshold function. W.l.o.g. we may consider a representation $f(x) = \text{sign}(w_0 + \sum_{i=1}^n w_i x_i)$ in which each $w_i \neq 0$, and by scaling the weights we may further assume that $T = [n] \setminus H$ has $\sum_{i \in T} w_i^2 = 1$.

We apply Theorem 11 to $f$. Parseval's identity implies that at most $1/\kappa(\epsilon)^4$ many indices $i$ can have $|\widehat{f}(i)| \geq \kappa(\epsilon)^2$, so we have $|H| \leq 1/\kappa(\epsilon)^4 = \text{poly}(1/\epsilon)$. In Case (i) we immediately have that $f$ is $O(\epsilon)$-close to a $\text{poly}(1/\epsilon)$-junta, so we suppose that Case (ii) holds, and henceforth argue about the $O(\epsilon)$-approximator $f'$ defined in Case (ii).

We consider all $2^{\text{poly}(1/\epsilon)}$ restrictions $\rho$ obtained by fixing the head variables in $H$. Our goal is to apply the results of Section 2.2 to the functions $f'|_\rho$. As noted in Section 2.3, for each restriction $\rho$ the resulting function $f'|_\rho$ over the tail variables in $T$ is a $\tau(\epsilon)$-regular threshold function, where $\tau(\epsilon) = O(\epsilon^{286})$ is the function implicit in the RHS of (3) (for brevity we henceforth write $\tau$ for $\tau(\epsilon)$). Moreover, all these restrictions are threshold functions defined by the same linear form over the variables in $T$: they only differ in their threshold values, i.e. the values $\theta_\rho \stackrel{\text{def}}{=} w_0 + \sum_{i \in H} w_i \rho_i$.

---

[3] See the discussion immediately before Equation (24) of [OS08]; our $\kappa(\epsilon)$ is the $\tau(\epsilon)$ of [OS08].

[4] See Equation (24) of [OS08].

In keeping with the notation of Section 2.2, for each restriction $\rho$ we write $h_{\theta_\rho}$ to denote $f'|_\rho$, i.e. $h_{\theta_\rho}(x_T) \stackrel{\text{def}}{=} \text{sign}(\theta_\rho + \sum_{i \in T} u_i x_i)$ where $u_i \stackrel{\text{def}}{=} \frac{\widehat{f}(i)}{\sigma_T}$ and $x_T \stackrel{\text{def}}{=} (x_i)_{i \in T}$. We observe that

$$\|u\|_1 = \sum_{i \in T} |u_i| = \frac{1}{\sigma_T} \sum_{i \in T} |\widehat{f}(i)| \leq \frac{1}{\sigma_T} \sum_{i=1}^n |\widehat{f}(i)| \leq \mathbb{I} \cdot \text{poly}(1/\epsilon). \tag{4}$$

where the last inequality uses $\text{Inf}_i(f) = |\widehat{f}(i)|$ and $\sigma_T = \Omega(\epsilon^2)$. Recalling that $N$ equals $\Theta(\|u\|_1^2 \cdot \frac{1}{\tau^2} \cdot \ln(1/\tau))$, we have that $N$ is at most $\mathbb{I}^2 \cdot \text{poly}(1/\epsilon)$.

We consider a distribution $\mathcal{D}''$ over threshold functions on $\{-1,1\}^n$ defined as follows: a draw of $g \leftarrow \mathcal{D}''$ is obtained by drawing $L \leftarrow \mathcal{D}$ and setting $g(x) = \text{sign}(w_0 + \sum_{i \in H} w_i x_i + \frac{\|u\|_1}{N} \cdot L(x_T))$. For every outcome of $g \leftarrow \mathcal{D}''$, the function $g$ depends on at most $|H| + N = (1 + \mathbb{I}^2) \text{poly}(1/\epsilon)$ many variables.

It remains only to argue that some $g$ drawn from $\mathcal{D}''$ is is $O(\epsilon)$-close to $f'$. Via the probabilistic method, to do this it suffices to show that $\mathbf{E}_{g \leftarrow \mathcal{D}''}[\mathbf{Pr}_{x \in \{-1,1\}^n}[g(x) \neq f'(x)]] = O(\tau)$ (recall that $\tau \ll \epsilon$). We now do this using the results of Section 2.2.

Fix any assignment $\rho$ to the variables in $H$. By Lemma 10 we have

$$\mathbf{E}_{g \leftarrow \mathcal{D}''}\left[\mathbf{Pr}_{x_T \leftarrow \{-1,1\}^{|T|}}[f'|_\rho(x_T) \neq g|_\rho(x_T)]\right] \leq 5\tau.$$

Averaging over all $\rho$, we get

$$\mathbf{E}_{g \leftarrow \mathcal{D}''}\left[\mathbf{Pr}_{x \leftarrow \{-1,1\}^n}[f'(x) \neq g(x)]\right] \leq 5\tau$$

which is the desired bound.

So, we have shown that every threshold function $f$ is $O(\epsilon)$-close to a $(1 + \mathbb{I}^2) \cdot \text{poly}(1/\epsilon)$-junta; we finish the proof of Theorem 1 by arguing that this implies a $\mathbb{I}^2 \cdot \text{poly}(1/\epsilon)$ junta size bound. Let $c$ be an absolute constant such that every $f$ is $\epsilon$-close to a $(1 + \mathbb{I}^2) \cdot (1/\epsilon)^c$-junta; we consider different cases based on the size of $\mathbb{I}$. If $\mathbb{I} > 1$, then it is clear that $(1 + \mathbb{I}^2)(1/\epsilon)^c < 2\mathbb{I}^2(1/\epsilon)^c < \mathbb{I}^2(1/\epsilon)^{c+1}$ (using $\epsilon < 1/2$). If $\mathbb{I} < \epsilon^2$, since $\sum_{|S| \geq 1} \widehat{f}(S)^2 \leq \sum_{|S| \geq 1} |S| \widehat{f}(S)^2 = \mathbb{I}$ (see Section 2.1.2), by Parseval's identity we get that $|\widehat{f}(\emptyset)| \geq 1 - \epsilon$. This means that $f$ is $\epsilon$-close to a constant function, which is of course a 0-junta. Finally, if $\epsilon^2 \leq \mathbb{I} \leq 1$, then $1 + \mathbb{I}^2 \leq 2 \leq 2\mathbb{I}^2 \epsilon^{-4} \leq \mathbb{I}^2 \epsilon^{-5}$, so $f$ can be $\epsilon$-approximated by a $\mathbb{I}^2(1/\epsilon)^{c+5}$-junta. So in every case $f$ is $\epsilon$-close to an $\text{Inf}(f)^2 \cdot (1/\epsilon)^{c+5}$-junta, and Theorem 1 is proved. $\square$

## 2.5 Discussion and Conjectures.

### 2.5.1 Improved low-weight approximators of threshold functions.

Recall the main result of [Ser07]:

**Theorem 12.** *[Ser07] Every $n$-variable threshold function $f$ is $\epsilon$-approximated by a threshold function $g = \text{sign}(w \cdot x - \theta)$ with $w_1, \ldots, w_n$ all integers satisfying $\sum_{i=1}^n w_i^2 \leq n \cdot 2^{\tilde{O}(1/\epsilon^2)}$.*

While a linear dependence on $n$ is the best possible bound which can hold uniformly for all $n$-variable threshold functions, it is possible to give a sharper bound that depends on $f$. Applying Theorem 12 to the threshold function junta which is given by Theorem 1, we obtain:

**Corollary 13.** *Every $n$-variable threshold function $f$ is $\epsilon$-approximated by a threshold function $g = \text{sign}(w \cdot x - \theta)$ with $w_1, \ldots, w_n$ all integers satisfying $\sum_{i=1}^{n} w_i^2 \leq \text{Inf}(f)^2 \cdot 2^{\tilde{O}(1/\epsilon^2)}$.*

Since $\text{Inf}(f)^2$ is at most $n$ (but can be much less) for every threshold function $f$, this strengthens Theorem 12.

### 2.5.2 A lower bound.

We observe that the $\text{Inf}(f)^2 \cdot \text{poly}(1/\epsilon)$ upper bound of Theorem 1 is nearly best possible: no strengthening can replace this with a bound smaller than $\Omega(\text{Inf}(f)^2 + 1/\epsilon^2)$.

For the $\Omega(\text{Inf}(f)^2)$ term, straightforward probability arguments (see e.g. [Ser07]) show that any $1/10$-approximator for the majority function $\text{sign}(x_1 + \cdots + x_n)$ must depend on $\Omega(n)$ variables. Since the total influence of majority is $\Theta(\sqrt{n})$, this shows that no subquadratic dependence on $\text{Inf}(f)$ is possible.

For the $\Omega(1/\epsilon^2)$ term, we use the following:

**Proposition 14.** *There is a threshold function $f$ with $\text{Inf}(f) = \Theta(1)$ such that any $\epsilon$-approximator $g$ must depend on $\Omega(1/\epsilon^2)$ variables.*

*Proof.* Let $a = \log(1/\epsilon) - 5$ and $b = 1/\epsilon^2$. The desired $f$ is

$$f(x) = \text{sign}(x_1 + \cdots + x_a + \frac{1}{2b}x_{a+1} + \cdots + \frac{1}{2b}x_{a+b} - a).$$

It can be verified that $\text{Inf}_i(f) = \Theta(\epsilon)$ for $i \in [a]$ and $\text{Inf}_i(f) = \Theta(\epsilon^2)$ for $i \in [a+1, b]$, so $\text{Inf}(f) = \Theta(1)$. Any $\epsilon$-approximator for $f$ must be a $1/16$-approximator of the subfunction $f|_\rho$ obtained by setting all the first $a$ bits to 1. But $f|_\rho$ is the majority function over $b$ variables, and as mentioned above any $1/16$-approximator must depend on $\Omega(b)$ variables. □

### 2.5.3 Extending to degree-$d$?

It is natural to wonder whether Theorem 1 extends to *polynomial* threshold functions (PTFs) of degree $d$, i.e. Boolean functions $f(x) = \text{sign}(p(x))$ where $p$ is a degree-$d$ polynomial. We pose the following conjecture which is a broad generalization of Theorem 1:

**Conjecture 1.** *Every degree-$d$ PTF $f$ is $\epsilon$-approximated by a $(\text{Inf}(f)/\epsilon)^{O(d)}$-junta.*

We suspect that even the $d = 2$ case of Conjecture 1 may be challenging, as the total influence of low-degree polynomial threshold functions does not seem to be well understood.

### 2.5.4 An exponential sharpening of Bourgain's theorem?

Recall that by Parseval's identity, every Boolean function $f$ has $\sum_{S \subseteq [n]} \hat{f}(S)^2 = 1$. Since the total influence $\mathrm{Inf}(f)$ equals $\sum_S \hat{f}(S)^2 |S|$ and the degree of each monomial $x_S$ is $|S|$, we may interpret $\mathrm{Inf}(f)$ as the "average" Fourier degree of $f$.

With this point of view, Friedgut's theorem may be viewed as part of a sequence of three results, all of which essentially say that Boolean functions with low degree (in some sense) are close (in some sense) to juntas. The first and earliest of these results is the following theorem of Nisan and Szegedy:

**Theorem 15.** *[NS94] Every Boolean function with (maximum) Fourier degree $k$ is a $k2^k$-junta.*

This theorem imposes a strong degree condition on $f$ – that it have *zero* Fourier weight above degree $k$ – and gets a strong conclusion, that $f$ is *identical* to a $k2^k$-junta. Next, Friedgut's theorem [Fri98] relaxed both the degree condition on $f$ and the resulting conclusion: if the "average" Fourier degree of $f$ (i.e. $\mathrm{Inf}(f)$) is at most $k$, then $f$ is $\epsilon$-close to a $2^{O(k/\epsilon)}$-junta. Finally and most recently, Bourgain relaxed the degree condition even further, by showing that if $f$ puts most of its Fourier weight on low-degree monomials, then regardless of where the remaining Fourier weight lies, $f$ must be close to a junta:

**Theorem 16.** *[Bou02] Every Boolean function $f$ with $\sum_{|S|>k} \hat{f}(S)^2 \leq (\epsilon/k)^{1/2+o(1)}$ is $\epsilon$-close to a $2^{O(k)} \cdot \mathrm{poly}(1/\epsilon)$-junta.*

Let us consider how each junta size bound changes when we restrict our attention to threshold functions in the above theorems. We first observe that the [NS94] bound can be exponentially improved in this case:

**Proposition 17.** *Every threshold function with (maximum) Fourier degree $k$ is a $(2k-1)$-junta.*

(This follows from the easy fact that any threshold function with $r$ relevant variables contains a subfunction which is an $(\frac{r+1}{2})$-way AND or OR.) Our Theorem 1, of course, tells us that Friedgut's theorem can also be exponentially sharpened if $f$ is a threshold function. This motivates the natural question of whether Bourgain's theorem can be similarly sharpened for threshold functions. We state the following:

**Conjecture 2.** *Every threshold function $f$ with $\sum_{|S|>k} \hat{f}(S)^2 \leq (\epsilon/k)^{1/2+o(1)}$ is $\epsilon$-close to a $\mathrm{poly}(k/\epsilon)$-junta.*

## 3 Theorem 2: approximating threshold functions to higher accuracy.

As outlined in Section 1.2, our new approach can be conceptually broken into the following steps:

1. Show that every threshold function has a representation in which many weights are "nice".

2. Use the "niceness" of the weights to establish anti-concentration of $w \cdot x$.

3. Finally, use the anti-concentration of $w \cdot x$ to obtain an approximator with small integer weights.

Note that there is a delicate relationship between the first two steps: the structural result for the weights that is established in the first step must match the necessary conditions for anti-concentration in the second step. The third step is a simple generic lemma translating anti-concentration into low-weight approximation.

The structure of this section is as follows: In Section 3.1 we recall the anti-concentration results that we need to implement Step 2 in our above proof template and prove the simple lemma that implements Step 3 in our proof template. In Section 3.2 we give a "warmup" to our main result by using the template to give a clean and modular proof of the main result of [Ser07]. In Section 3.3 we show how the template yields a variant of Theorem 2 which has an $n^{O(1/\epsilon^{2/3})}$ bound. This subsection includes the main new technical contribution of Section 3, a new result on representations of threshold functions, Lemma 26. Roughly speaking, this lemma says that every threshold function has a representation such that many of the differences between consecutive weights are not too small. Then in Section 3.4 we show how this $n^{O(1/\epsilon^{2/3})}$ bound can be improved to fully prove Theorem 2.

Finally, all the results of this section can be appropriately generalized to constant-biased product distributions and $K$-wise independent distributions (but as we show, they provably *cannot* be generalized to *every* distribution). We give these extensions in Section 4.

## 3.1 Anti-concentration of weighted sums of Bernoulli random variables.

We start with the formal definition of anti-concentration:

**Definition 18.** *Let $a \in \mathbb{R}^n$ be a weight–vector and $r \in \mathbb{R}_+$. The* Lévy *anti-concentration function of $a$ is defined as*

$$p_r(a) \stackrel{def}{=} \sup_{v \in \mathbb{R}} \mathbf{Pr}_{x \leftarrow \mathcal{U}}[|a \cdot x - v| \leq r].$$

Thus, the anti-concentration of a weight vector $a$ is an upper bound on the probability that $a \cdot x$ lies in any small interval (of length $2r$). An early and important result on anti-concentration was given by Erdős [Erd45]; improving on an earlier result of Littlewood and Offord [LO43], he proved

**Theorem 19** (Erdős). *Let $a = (a_1, \ldots, a_k) \in \mathbb{R}^k$, $r \in \mathbb{R}_+$ be such that $|a_i| \geq r$ for all $i \in [k]$. Then $p_r(a) \leq \binom{k}{k/2}/2^k = O(k^{-1/2})$.*

A large body of subsequent work generalized this result in many different ways (see e.g. Chapter 7 of [TV06]); anti-concentration results of this general flavor have come to be known as "Littlewood-Offord theorems." We shall require an extension of Theorem 19 which is due to Halász [Hal77], improving upon Erdős-Moser [Erd65] and Sárközy-Szemerédi [SS65]. While Erdős's theorem gives the best (smallest) possible anti-concentration bound assuming that each *weight* $a_i$ is large, Halász's theorem gives a stronger bound under the stronger assumption that the *difference* between any two weights is large:

**Theorem 20** (Halász). *Let $a = (a_1, \ldots, a_k) \in \mathbb{R}^k$, $r \in \mathbb{R}_+$ be such that $|a_i - a_j| \geq r$ for all $i \neq j \in [k]$. Then $p_r(a) \leq O(k^{-3/2})$.*

Looking ahead, we note that the "3/2" exponent instead of "1/2" is the key to our improvement from $2^{\tilde{O}(1/\epsilon^2)}$ to $2^{\tilde{O}(1/\epsilon^{2/3})}$.

The last fact about anti-concentration that we shall need is the following simple lemma, which says that if we extend a weight vector $a$ by adding more weights, its anti-concentration can only improve:

**Lemma 21** (Extension). *Let $a \in \mathbb{R}^k$ be any $k$-dimensional weight vector and $r \in \mathbb{R}_+$ be any non-negative real. For any $n > k$, let $a' \in \mathbb{R}^n$ be the vector $(a_1, \ldots, a_k, a'_{k+1}, \ldots, a'_n)$ where the weights $a'_{k+1}, \ldots, a'_n$ may be any real numbers. Then we have $p_r(a') \leq p_r(a)$.*

The proof is by a simple averaging argument, using the fact that for $x \leftarrow \{-1, 1\}^n$ uniform random, conditioned on any outcome of the variables $x_{k+1}, \ldots, x_n$, the distribution of $x_1, \ldots, x_k$ is still uniform.

**From anti-concentration to a low-weight approximator.** The following simple lemma takes us from anti-concentration to a low-weight approximator. We use it to implement Step 3 in our proof template.

**Lemma 22.** *Let $g = \text{sign}(\sum_{i=1}^n w_i x_i - \theta)$ be any threshold function. If $p_r(w_1, \ldots, w_n) \leq \epsilon$, then there exists a $2\epsilon$-approximator $h$ for $g$, where $h$ is a threshold function with integer weights each of magnitude $O(\max_i |w_i| \cdot \sqrt{n \ln(1/\epsilon)}/r)$.*

*Proof.* Let $\alpha = r/(\sqrt{n \ln(2/\epsilon)})$. For each $i \in [n]$, let $u_i$ be the value obtained by rounding $w_i$ to the nearest integer multiple of $\alpha$ and $v_i = u_i/\alpha \in \mathbb{Z}$. We claim that $h(x) = \text{sign}(\sum_{i=1}^n v_i x_i - \theta/\alpha)$ is the desired approximator. It is clear that $\max_i |v_i| = O(\max_i |w_i|/\alpha)$, so it suffices to show that $h$ is $(\epsilon + \epsilon)$-close to $g$.

For $i \in [n]$, let $e_i = w_i - u_i$, so that $u \cdot x = w \cdot x - e \cdot x$. We have that $g(x) \neq h(x)$ only if $|e \cdot x| \geq r$ or $|w \cdot x - \theta| \leq r$. We bound from above the probability of each of these events by $\epsilon$. The probability of the second event is bounded by $\epsilon$ since $\mathbf{Pr}[|w \cdot x - \theta| \leq r] \leq p_r(w) \leq \epsilon$. For the first event we have $\mathbf{Pr}[|e \cdot x| \geq r] \leq \mathbf{Pr}[|e \cdot x| \geq \|e\|_2 \sqrt{2 \ln(2/\epsilon)}] \leq \epsilon$, where the first inequality uses the fact $\|e\|_2 \leq (r/\sqrt{2 \ln(2/\epsilon)})$ and the second follows from the Hoeffding bound. □

## 3.2  Warmup: Simple Proof of [Ser07] Main Result.

In this section we give a simple and modular proof of nearly the same bound as the main result of [Ser07], following the proof template from the start of Section 3. Let $f : \{-1, 1\}^n \to \{-1, 1\}$ be any threshold function.

**First step:** This is provided for us by the following result, which is an immediate consequence of Lemma 14 in [OS08]. Intuitively, this result says that every threshold function has a representation in which the $k$-th largest weight is not too small compared with the largest weight.[5]

**Claim 23.** *Let $f : \{-1, 1\}^n \to \{-1, 1\}$ be any threshold function, let $\epsilon > 0$, and let $k \in [n]$. There is an $\epsilon$-approximator $g = \text{sign}(\sum_{i=1}^n w_i x_i - \theta)$ for $f$ with the following property: Suppose (reordering and rescaling weights if necessary) that $1 = |w_1| \geq \cdots \geq |w_n|$. Then $|w_k| \geq 1/(k^k \sqrt{3n \ln(2/\epsilon)})$.*

**Second step:** We apply Erdős's theorem, Theorem 19, to the weight vector $(w_1, \ldots, w_k)$ from Claim 23 (we will fix $k$ later), taking $r = 1/(k^k \sqrt{3n \ln(2/\epsilon)})$ to be the bound from Claim 23. Theorem 19 gives $p_r(w_1, \ldots, w_k) \leq O(1/\sqrt{k})$, and the Extension Lemma 21 gives that in fact $p_r(w_1, \ldots, w_n) \leq O(1/\sqrt{k})$.

**Third step:** It remains only to fix $k = \min\{1/\epsilon^2, n\}$ and observe that the $h$ obtained from Lemma 22 is an $O(\epsilon)$-approximator for $f$. (Note that if $1/\epsilon^2 > n$, then integer weights $2^{\tilde{O}(1/\epsilon^2)}$ suffice to *exactly* represent $f$ by [MTT61].) We have thus proved:

**Theorem 24.** *Every $n$-variable threshold function $f$ is $\epsilon$-approximated by a threshold function $h = \text{sign}(v \cdot x - \theta)$ with $v_1, \ldots, v_n$ all integers of magnitude $n \cdot 2^{\tilde{O}(1/\epsilon^2)}$.*

This is almost identical to the main result of [Ser07]; the bound of [Ser07] is slightly stronger (it has $\sqrt{n}$ in place of $n$).

## 3.3 Toward Theorem 2: An $n^{O(1/\epsilon^{2/3})}$ bound.

In this section we prove an intermediate result towards our ultimate goal of $\text{poly}(n) \cdot 2^{\tilde{O}(1/\epsilon^{2/3})}$:

**Theorem 25.** *Every $n$-variable threshold function $f$ is $\epsilon$-approximated by a threshold function $h = \text{sign}(v \cdot x - \theta)$ with $v_1, \ldots, v_n$ all integers of magnitude $n^{O(1/\epsilon^{2/3})}$.*

We follow the same high-level proof template as the previous section. Let $f : \{-1, 1\}^n \to \{-1, 1\}$ be a threshold function. We may assume w.l.o.g. that $f$ depends on all $n$ input variables, and since the claimed bound follows again from [MTT61] if $1/\epsilon^{2/3} > n - 2$, we assume $1/\epsilon^{2/3} \leq n - 2$.

**First step:** Our goal now is to apply Halász's anti-concentration bound in Step 2 rather than Erdős's theorem. To do this we need the following new result on representing threshold functions, which intuitively says that every threshold function has a representation using weights such that many of the differences between consecutive weights are not too small compared to the largest weight:

---
[5]We do not repeat the proof of Claim 23 or Lemma 14 from [OS08] here but we note that the proof is self-contained and rather straightforward; it follows along the lines of [MTT61]'s classic argument to upper bound the weights required to represent any threshold function.

**Lemma 26.** *Let $f : \{-1,1\}^n \to \{-1,1\}$ be a threshold function that depends on all $n$ variables. There is a representation $\mathrm{sign}(\sum_{i=1}^n w_i x_i - \theta)$ for $f$ with the following property: Suppose (reordering and rescaling weights if necessary) that $1 = |w_1| \geq \cdots \geq |w_n| > 0$. For $i \in [n-1]$ let $\Delta_i \overset{\text{def}}{=} |w_i| - |w_{i+1}|$. Then for any $k \in [n-2]$, the $k$-th biggest element of the (multiset) $\Delta_1, \ldots, \Delta_{n-1}$ is at least $\frac{1}{(2n+2)^{2k+8}}$.*

We pause to contrast this result with an earlier theorem due to Håstad [Hås05] that appeared in [Ser07]. Under the same hypotheses as Lemma 26, the earlier theorem asserted that for any $k \in [n]$ the $k$-th largest weight $w_k$ satisfies $|w_k| \geq \frac{1}{k!(n+1)}$. The proof of the earlier theorem centers on a careful analysis of a linear program in which the variables are the weights $w_1, \ldots, w_n$ and there are $2^n$ constraints corresponding to the $2^n$ points $x \in \{-1,1\}^n$. To prove Lemma 26, we must now analyze a linear program with some additional constraints which, intuitively, ensure that there are "gaps" between the weights.[6] We prove Lemma 26 at the end of this subsection.

**Second step:** We take $k = 1/\epsilon^{2/3}$ and consider the $k$ largest differences $\Delta_{i_1} = |w_{i_1}| - |w_{i_1+1}|, \ldots, \Delta_{i_k} = |w_{i_k}| - |w_{i_k+1}|$. Lemma 26 implies that for all $a \neq b \in [k]$ we have $|w_{i_a} - w_{i_b}| \geq r$, for $r = 1/(2n+2)^{2k+8}$. Applying Halász's anti-concentration bound, Theorem 20, we get that $p_r(w_{i_1}, \ldots, w_{i_k}) \leq O(k^{-3/2}) = O(\epsilon)$, and the Extension Lemma 21 further gives $p_r(w_1, \ldots, w_n) = O(\epsilon)$.

**Third step:.** We simply apply Lemma 22. Recalling that $r = 1/(2n+2)^{\Theta(1/\epsilon^{2/3})}$, the proof of Theorem 25 is complete (modulo the proof of Lemma 26). □

*Proof of Lemma 26.* Let $f(x)$ be a threshold function. We first consider the case that $f$ is odd, i.e. $f(x) = -f(-x)$ for all $x \in \{-1,1\}^n$; in this case $f$ can be represented with a threshold of zero. Once we have established the result for such threshold functions we will use it to establish the general case.

By symmetry of $\{-1,1\}^n$ we may assume that $f$ is monotone increasing in each coordinate $x_i$. By reordering coordinates we may assume that $\mathrm{Inf}_1(f) \geq \mathrm{Inf}_2(f) \geq \cdots \geq \mathrm{Inf}_n(f) > 0$ (the final inequality is strict because $f$ depends on all $n$ coordinates).

We consider the set $\mathcal{W} \subseteq \mathbb{R}^n$ of weight vectors $w = (w_1, \ldots, w_n)$ that satisfy the following properties:

1. $w \cdot x \geq 1$ for every $x \in \{-1,1\}^n$ such that $f(x) = 1$. Note that since $f$ is odd these inequalities imply the corresponding inequalities for negative points, $w \cdot x \leq -1$ for every $x \in \{-1,1\}^n$ such that $f(x) = -1$.

2. $w_i - w_{i+1} \geq 1$ for all $i = 1, 2, \ldots, n-1$, and $w_n \geq 1$.

The first set of $2^{n-1}$ constraints says that $\mathrm{sign}(w \cdot x)$ is a valid representation for $f$ (i.e. $f(x) = \mathrm{sign}(w \cdot x)$ for all $x \in \{-1,1\}^n$). The second set of $n$ constraints says that no two weights are precisely the same and moreover all the weights are positive. (These are the new constraints that did not feature in the proof of [Hås05].)

---

[6]In fact, by considering the majority function one can verify that the $2^n$-constraint linear program of the earlier proof is not sufficient; that LP yields a representation in which each $w_i$ is the same and hence the "gaps" $\Delta_i$ are all 0.

Thus $\mathcal{W}$ is the feasible set of a linear program $\mathcal{LP}$ consisting of $2^{n-1} + n$ inequalities on $w_1, \ldots, w_n$: $2^{n-1}$ inequalities correspond to points of the hypercube $\{-1, 1\}^n$ and $n$ inequalities correspond to the set

$$D_n = \{(1, -1, 0, \ldots, 0)_{1 \times n}, (0, 1, -1, 0, \ldots, 0)_{1 \times n}, \ldots, (0, \ldots, 1, -1)_{1 \times n}, (0, \ldots, 0, 1)_{1 \times n}\}.$$

We claim that the linear program $\mathcal{LP}$ is feasible, or equivalently $\mathcal{W} \neq \emptyset$. Indeed, by simple standard arguments it can be shown that every odd threshold function $f : \{-1, 1\}^n \to \{-1, 1\}$ has a representation $\text{sign}(w \cdot x)$ such that (i) for all $x \in \{-1, 1\}^n$, it holds $\text{sign}(w \cdot x) \neq 0$, and (ii) every partial sum of the weights is distinct, i.e. for all $I \neq J \subseteq [n]$ it holds $\sum_{i \in I} w_i \neq \sum_{i \in J} w_j$. The latter in particular implies that $w_1 \neq w_2 \neq \ldots \neq w_n$. Now recall that $\text{Inf}_1(f) \geq \text{Inf}_2(f) \geq \ldots \geq \text{Inf}_n(f) > 0$ and that $f$ is monotone increasing in all its coordinates. It is well known and easy to show (see e.g. [FP04]) that there is a representation $\text{sign}(w \cdot x)$ of such a threshold function that satisfies $w_1 \geq w_2 \geq \ldots w_n > 0$. Therefore, we can scale the weights so that all the constraints in the linear program $\mathcal{LP}$ are simultaneously satisfied.

Having established that $\mathcal{W} \neq \emptyset$, we select a weight vector $w^* \in \mathcal{W}$ that maximizes the number of *tight* inequalities (i.e. satisfied with equality) in $\mathcal{LP}$. If more than one weight vector satisfies a maximum number of tight inequalities, we choose one arbitrarily. At this point, we invoke the following crucial claim:

**Claim 27.** *There exists a set of $n$ points $y^{(1)}, \ldots, y^{(n)} \in f^{-1}(1) \cup D_n$ such that $w^*$ is the unique solution of the linear system: $\{w \cdot y^{(i)} = 1 \mid i = 1, 2, \ldots, n\}$. (Henceforth, we shall denote this system by $(*)$.)*

The proof of the claim is essentially the same as in the proof of Muroga *et al.*'s [MTT61] classic upper bound on the size of integer weights that are required to express LTF's over $\{-1, 1\}^n$. For completeness we include a proof of the claim in Appendix 5.

Note that $(*)$ is a system of $n$ linear equations in the variables $w_1, \ldots, w_n$ where each coefficient of each variable in the equations is $-1, 0$ or $1$ and the right-hand side of each equation is 1. Since our goal is to prove a statement about the magnitude of the differences $w_i - w_{i+1}$, $i = 1, 2, \ldots, n-1$, we define an appropriate set of $n$ new variables and rewrite $(*)$. In particular, we define the set of variables $\delta_1, \ldots, \delta_n$ as follows:

$$\delta_n = w_n, \qquad \delta_i = w_i - w_{i+1} \text{ for } i = 1, \ldots, n-1.$$

This is equivalent to

$$w_n = \delta_n, \qquad w_i = \delta_i + \cdots + \delta_n \text{ for } i = 1, \ldots, n-1.$$

We let $\delta$ denote $[\delta_1, \ldots, \delta_n]$. By rewriting $(*)$, we get an *equivalent* system $(**)$ of $n$ equations in variables $\delta_1, \ldots, \delta_n$ where the coefficients of each variable in each equation are integers in the range $[-n, n]$ and all the right-hand sides remain 1. Hence, the linear system $(**)$ has the unique *strictly* positive solution

$$\delta_n^* = w_n^*, \qquad \delta_i^* = w_i^* - w_{i+1}^* \text{ for } i = 1, \ldots, n-1.$$

At this point we reorder the variables $\delta_i$ in decreasing order of magnitude of the $\delta_i^*$'s. We thus get a new set of variables $\tau_1, \ldots, \tau_n$ such that

$$\tau_i^* = i\text{-th largest of } \{\delta_1^*, \ldots, \delta_n^*\},$$

breaking ties arbitrarily. We similarly denote $\tau = [\tau_1, \ldots, \tau_n]$.

So $(**)$ is now a system of $n$ equations in variables $\{\tau_i\}_{i \in [n]}$, where the coefficients of each variable in each equation are integers in the range $[-n, n]$ and all the right-hand sides are still 1. The values $\tau_1^*, \ldots, \tau_n^*$ in the unique solution of this system are strictly positive and ordered in decreasing order of magnitude. Let us write

$$\alpha_{j1}\tau_1 + \alpha_{j2}\tau_2 + \ldots + \alpha_{jn}\tau_n = 1$$

for the $j$-th equation where $\alpha_{ij}$, $i, j \in [n]$ are integers in $[-n, n]$. It is not difficult to see that the above system is equivalent to the following system of $n$ equations in $\tau_1, \ldots, \tau_n$:

$$\alpha_{j1}\tau_1 + \alpha_{j2}\tau_2 + \ldots + \alpha_{jn}\tau_n = \alpha_{11}\tau_1 + \alpha_{12}\tau_2 + \ldots + \alpha_{1n}\tau_n \text{ for } j = 2, 3, \ldots, n, \quad \text{and} \quad \tau_n = \tau_n^*.$$

Each of the first $n-1$ equations is homogeneous and can be rewritten as $\tau \cdot z^{(j)} = 0$, where $z^{(j)}$ is a vector whose entries are integers in $[-2n, 2n]$. So we have that $\tau^* = [\tau_1^*, \ldots, \tau_n^*]$ is the unique solution to a linear system:

$$Z\tau = b \tag{5}$$

where $Z$ is a non-singular $n \times n$ matrix with entries that are integers in $[-2n, 2n]$ and with last row $(0, \ldots, 0, 1)$, and $b$ is $[0, 0, \ldots, 0, \tau_n^*]$.

Recall that $\tau_1^* \geq \cdots \geq \tau_n^* > 0$. We now show that each $\tau_k^*$ is somewhat large compared to $w_1^*$. The case $k = 1$ is easy: since $\sum_{i=1}^n \tau_i^* = w_1^*$, we have $\tau_1^* \geq w_1^*/n$.

Fix any $k \in \{2, \ldots, n\}$. After possibly reordering the rows of $Z$, the $(k-1)$-dimensional vector $[1, 0, \ldots, 0]$ can be expressed as a linear combination $a_1 R_1 + \cdots + a_{k-1} R_{k-1}$ where $R_i$ is the $i$-th row of the $(k-1) \times (k-1)$ upper left submatrix of $Z$. Since all entries in $Z$ are integers in $[-2n, 2n]$, Cramer's Rule implies that each $|a_i|$ is at most the maximum determinant of any $(k-1) \times (k-1)$ matrix with all entries in $[-2n, 2n]$; this is easily seen to be at most $(k-1)!(2n)^{k-1}$. It follows that there is a linear combination of the first $k-1$ equations of (5) which yields

$$\tau_1 = \sum_{j=k}^n \gamma_j^k \tau_j \tag{6}$$

where each $|\gamma_j^k|$ is at most $(k-1) \cdot (2n) \cdot (k-1)!(2n)^{k-1} \leq (2(k-1)n)^k$. From (6), setting $\tau = \tau^*$ and recalling that the $\tau_i^*$'s are positive and ordered by magnitude, we now get $\tau_1^* \leq (n-k+1)(\max_j |\gamma_j^k|)\tau_k^*$ which implies

$$\tau_k^* \geq \frac{\tau_1^*}{(2(k-1)n)^k(n-k+1)} \geq \frac{\tau_1^*}{(2n)^{2k+1}} \tag{7}$$

Observing that $\sum_{i=1}^n \tau_i^* = w_1^*$, we have $\tau_1^* \geq w_1^*/n$, which implies

$$\tau_k^* \geq \frac{w_1^*}{(2n)^{2k+2}}.$$

Finally, we observe that for $k \in [n-1]$, the $k$-th biggest element of the multiset $\Delta_1, \ldots, \Delta_{n-1}$ (see the lemma statement) is at least $\tau_{k+1}$. (It is either $\tau_{k+1}$ or $\tau_k$ depending on whether or not $\delta_n^* = w_n^*$ is among the $k$ largest elements of $\{\delta_1^*, \ldots, \delta_n^*\}$.) Renormalizing so that the largest weight is 1, we have shown that for odd $f$, the $k$-th biggest element of the multiset $\Delta_1, \ldots, \Delta_{n-1}$ is at least $\frac{1}{(2n)^{2k+4}}$. This completes the proof of Lemma 26 for the case that $f$ is odd.

We now treat the case where $f$ is not odd, i.e. $f$ has a nonzero threshold. We do this by considering the threshold function $g : \{-1,1\}^{n+1} \to \{-1,1\}$ which has zero threshold, $n$ weights the same as $f$, and an $(n+1)$-st weight which is the threshold of $f$. The result for the zero-threshold case shows that $g$ has a representation $\text{sign}(w_1 x_1 + \cdots + w_n x_n + w_{n+1} x_{n+1})$ where $|w_1| \geq \cdots \geq |w_{n+1}|$, and letting $\Delta_i = |w_i| - |w_{i+1}|$ for $i \in [n]$, the $k$-th biggest element of $\Delta_1, \ldots, \Delta_n$ is at least $\frac{|w_1|}{(2n+2)^{2k+4}}$ for any $k \in [n]$.

We now observe that for $k \in [n-2]$, the $k$-th biggest gap between the magnitudes of the $w_i$'s that correspond to actual weights of $f$ is at least the $(k+2)$-th biggest element of $\Delta_1, \ldots, \Delta_n$. This holds since at most two of the values $\Delta_j = |w_j| - |w_{j+1}|$ can involve the weight $w_{j^*}$ which corresponds to the threshold of $f$, as opposed to one of its actual weights. Since $|w_1|$ is at least as large as the absolute value of the largest actual weight of $f$, we get that for $k \in [n-2]$, the $k$-th biggest gap between the magnitudes of the actual weights of $f$ is at least (largest weight of $f$)/$(2n+2)^{2k+8}$. Renormalizing so that the largest magnitude weight of $f$ is 1, Lemma 26 is proved. □

## 3.4 Proof of Theorem 2: A $\text{poly}(n) \cdot 2^{\tilde{O}(1/\epsilon^{2/3})}$ bound.

Given a threshold function $f(x) = \text{sign}(w \cdot x - \theta)$ such that $|w_1| \geq \cdots \geq |w_n| > 0$, for $k \in [n]$ we denote by $\sigma_k$ the quantity $\sqrt{\sum_{i=k}^n w_i^2}$. The analysis in [Ser07] is based on the notion of the "$\tau$-critical index":

**Definition 28.** *We define the $\tau$-critical index $\ell(\tau)$ of a threshold function $f = \text{sign}(w \cdot x - \theta)$ as the smallest index $i \in [n]$ for which $|w_i| \leq \tau \cdot \sigma_i$. If this inequality does not hold for any $i \in [n]$, we define $\ell(\tau) = \infty$.*

We now show how to use Theorem 25 and ideas from [Ser07] to prove Theorem 2. Given $\epsilon > 0$, we proceed by a case analysis, as in [Ser07], based on the value of the $\epsilon$-critical index $\ell \stackrel{\text{def}}{=} \ell(\epsilon)$. If $\ell > L \stackrel{\text{def}}{=} \tilde{O}(1/\epsilon^2)$, Case IIa in [Ser07] says that $f$ is $\epsilon$-close to the $L$-junta $g$ obtained by truncating the smallest $(n-L)$ weights, i.e. $g(x) = \text{sign}(\sum_{i=1}^L w_i x_i - \theta)$. By applying Theorem 25 to $g$, we obtain an $\epsilon$-approximator $h$ with integer weights of magnitude $L^{O(1/\epsilon^{2/3})} = 2^{\tilde{O}(1/\epsilon^{2/3})}$, which is a $2\epsilon$-approximator for $f$. It remains to handle the case $\ell \leq L$. To do this, we use another fact from [Ser07]; that, for every value of $\ell$, there exists an $\epsilon$-approximator for $f$ with integer weights of magnitude $\sqrt{n \ln(1/\epsilon)} \cdot 2^{O(\ell \log \ell)}$. If $\ell < K \stackrel{\text{def}}{=} 2/\epsilon^{2/3}$, this yields an $\epsilon$-approximator with integer weights of magnitude $\sqrt{n} \cdot 2^{\tilde{O}(1/\epsilon^{2/3})}$ and we are done. To handle the case $K \leq \ell \leq L$, we use a combination of Gaussian anti-concentration (for the $n - \ell + 1$ smallest weights) and "Halász-type" anti-concentration (for the largest $\ell - 1$ weights).

Let us proceed with the analysis. We start by rounding the weights $w_\ell, \ldots, w_n$, exactly as in Case IIb in [Ser07], to get an $\epsilon$-approximator $g(x) = \text{sign}(\sum_{i=1}^n v_i x_i - \theta')$ for $f$ with the following properties: (i) For $i \geq \ell$, each $v_i$ is an integer of magnitude $O(\sqrt{n \ln(1/\epsilon)})$ and $\sum_{i=\ell}^{n} v_i^2 = O(n \ln(1/\epsilon)/\epsilon^2)$; (ii) It holds $|v_1| \geq |v_2| \geq \ldots \geq |v_{\ell-1}| > 1$. Our goal is to establish the existence of an $\epsilon$-approximation $h$ for $g$ with small integer weights. To achieve this, we will use the fact that the "tail" of $g$ has small integer coefficients, i.e. the integer–valued random variable $t(x) \stackrel{\text{def}}{=} \sum_{i=\ell}^{n} v_i x_i$ has small support.

Let $R, k > 0$ be integers. Denote by $\Omega(R, k)$ the set $\{\pm 1\}^{k-1} \times \{-R, -R+1, \ldots, R-1, R\}$. Now fix an integer $R_0 = \Theta(\sqrt{n} \ln(1/\epsilon)/\epsilon)$ and denote $\Omega_0 \stackrel{\text{def}}{=} \Omega(R_0, \ell)$. Consider the threshold function $h : \Omega_0 \to \{\pm 1\}$ defined by $h(y) = \text{sign}(\sum_{i=1}^{\ell-1} v_i y_i + y_\ell - \theta')$, $y \in \Omega_0$. We claim that the threshold function $g' : \{-1, 1\}^n \to \{-1, 1\}$ defined by $g'(x) = h(x_1, \ldots, x_{\ell-1}, t(x))$ is $\epsilon$-close to $g$. To see this note that $g'(x)$ equals $g(x)$ whenever $|t(x)| = |\sum_{i=\ell}^{n} v_i x_i| \leq R_0$, and this holds for a random $x$ with probability $1 - \epsilon$ by a Hoeffding bound (since $R_0 \geq \sqrt{2 \ln(2/\epsilon) \sum_{i=\ell}^{n} v_i^2}$ by the definition of $R_0$ and property (i) of $g$).

At this point we use the following technical generalization of Lemma 26, whose proof is deferred to the end of this subsection:

**Lemma 29.** *Let $h' : \Omega(R, k) \to \{\pm 1\}$ be a threshold function that depends on all $k$ variables. Suppose that $h'(y)$ has a representation as $\text{sign}(\sum_{i=1}^{k} w_i' y_i - \theta')$ such that $|w_1'| \geq |w_2'| \geq \ldots \geq |w_k'| > 0$. There exists an alternate representation of $h'$ as $\text{sign}(\sum_{i=1}^{k} u_i y_i - \theta'')$ satisfying $1 = |u_1| \geq \cdots \geq |u_k| > 0$, with the following property: For $i \in [k-1]$ let $\Delta_i \stackrel{\text{def}}{=} |u_i| - |u_{i+1}|$. Then for any $j \in [k-2]$, the $j$-th biggest element of the (multiset) $\Delta_1, \ldots, \Delta_{k-1}$ is at least $\frac{1}{(2k+2R)\cdot(2k+2)^{2j+8}}$.*

Applying this lemma to $h$, i.e. setting $h' = h$, $R = R_0$ and $k = \ell$, and fixing $j \stackrel{\text{def}}{=} 1/\epsilon^{2/3} + 2 < K - 2 \leq \ell - 2$, we obtain a representation $\text{sign}(\sum_{i=1}^{\ell} u_i y_i - \theta'')$ for $h$ such that the $j$ largest differences $\Delta_{i_1} = |u_{i_1}| - |u_{i_1+1}|, \ldots, \Delta_{i_j} = |u_{i_j}| - |u_{i_j+1}|$ are at least $r_0$, for $r_0 = \frac{1}{(2\ell+2R_0)\cdot(2\ell+2)^{2j+8}} = (1/\sqrt{n}) \cdot 2^{-\tilde{O}(1/\epsilon^{2/3})}$. (Note that the latter equality uses the fact that $\ell \leq L$.) This yields a set of $j' = 1/\epsilon^{2/3}$ weights $u_{l_1}, \ldots, u_{l_{j'}}$ – not including $u_\ell$ – whose absolute differences are at least $r_0$, i.e. for all $a \neq b \in [j']$, we have $|u_{l_a} - u_{l_b}| \geq r_0$.

We are now ready to use our proof template again. The alternate representation for $h$ from above and the definition of $g'$ imply that $g'(x)$ can be represented as $\text{sign}(\sum_{i=1}^{\ell-1} u_i x_i + \sum_{i=\ell}^{n} u_i' x_i - \theta'')$, where $u_i' \stackrel{\text{def}}{=} u_\ell v_i$, $\ell \leq i \leq n$. By Halász's bound, Theorem 20, applied to the weights $u_{l_1}, \ldots, u_{l_{j'}}$, and the Extension Lemma 21 as before, we conclude that $p_{r_0}(u_1, \ldots, u_n') = O(\epsilon)$. Finally, since the maximum weight in (the new representation for) $g'$ is $O(\sqrt{n \log(1/\epsilon)})$ (as follows from the fact that $|u_i| \leq 1$, $i \in [\ell]$, and property (i) of $g$), Lemma 22 implies the existence of an $O(\epsilon)$-approximator for $g'$ with integer weights each at most $n^{3/2} \cdot 2^{\tilde{O}(1/\epsilon^{2/3})}$. This concludes the proof of Theorem 2. □

*Proof of Lemma 29:* The proof is a technical extension of Lemma 26 taking into account the fact that the last variable of $h'$ has a non-boolean range. We consider the same linear program as in Lemma 26 and the analysis extends essentially by following the previous proof line-by-line. We therefore omit some details.

We similarly start with the case that $h'$ can be represented with $\theta' = 0$. Once we have established the result for such threshold functions the general case follows easily. By symmetry of $\Omega(R, k)$ we may assume that $h'$ is monotone increasing in each coordinate $y_i$.

The linear program $\mathcal{LP}$ is the set $\mathcal{W} \subseteq \mathbb{R}^k$ of weight vectors $w = (w_1, \ldots, w_k)$ with the following properties:

1. $w \cdot y \geq 1$ for every $y \in \Omega(R, k)$ such that $h'(y) = 1$.

2. $w_i - w_{i+1} \geq 1$ for all $i = 1, 2, \ldots, k-1$, and $w_k \geq 1$.

Now the $\mathcal{LP}$ consists of $(2R+1) \cdot 2^{k-1} + k$ inequalities: $(2R+1) \cdot 2^{k-1}$ inequalities correspond to points of $\Omega(R, k)$ and $k$ inequalities correspond to the set

$$D_k = \{(1, -1, 0, \ldots, 0)_{1 \times k}, (0, 1, -1, 0, \ldots, 0)_{1 \times k}, \ldots, (0, \ldots, 1, -1)_{1 \times k}, (0, \ldots, 0, 1)_{1 \times k}\}.$$

We claim that the linear program $\mathcal{LP}$ is feasible. This follows from the fact that $h'$ is assumed to have a representation as $\text{sign}(\sum_{i=1}^{k} w'_i y_i - \theta')$ such that $|w'_1| \geq |w'_2| \geq \ldots \geq |w'_k| > 0$. (Even if some of the $w'_i$'s are equal, we can slightly perturb them without changing the function. Then, we can scale everything up if necessary and obtain a feasible solution to the $\mathcal{LP}$.) Given that $\mathcal{W} \neq \emptyset$, we similarly select a weight vector $w^* \in \mathcal{W}$ that maximizes the number of *tight* inequalities in $\mathcal{LP}$ (breaking ties arbitrarily). We invoke the following claim which is proved virtually identical to Claim 27:

**Claim 30.** *There exists a set of $k$ points $y^{(1)}, \ldots, y^{(k)} \in (h')^{-1}(1) \cup D_k$ such that $w^*$ is the unique solution of the linear system: $\{w \cdot y^{(i)} = 1 \mid i = 1, 2, \ldots, k\}$. (Henceforth, we shall denote this system by $(*)$.)*

Note that $(*)$ is a system of $k$ linear equations in the variables $w_1, \ldots, w_k$ where each coefficient of the variables $w_1, \ldots, w_{k-1}$ in the equations is in $\{-1, 0, 1\}$, the coefficients of $w_k$ are in $\{-R, -R+1, \ldots, R-1, R\}$ and the right-hand side of each equation is 1. Now, following the analysis of Lemma 26 line-by-line, we obtain the final system $Z\tau = b$, where $Z$ is a non-singular $k \times k$ matrix with entries that are integers in $[-2k, 2k]$ with the exception of one column with entries in $[-2k - 2R, 2k + 2R]$ and with last row $(0, \ldots, 0, 1)$; similarly, $b$ equals $[0, 0, \ldots, 0, \tau_k^*]$.

This system has a unique solution $\tau^* = [\tau_1^*, \ldots, \tau_k^*]$, where $\tau_1^* \geq \cdots \geq \tau_k^* > 0$. We similarly show that each $\tau_j^*$ is somewhat large compared to $w_1^*$. Fix any $j \in \{2, \ldots, k\}$. After possibly reordering the rows of $Z$, the $(j-1)$-dimensional vector $[1, 0, \ldots, 0]$ can be expressed as a linear combination $a_1 R_1 + \cdots + a_{j-1} R_{j-1}$ where $R_i$ is the $i$-th row of the $(j-1) \times (j-1)$ upper left submatrix of $Z$. Since all entries in $Z$ are integers in $[-2k, 2k]$, except for one column with entries in $[-2k - 2R, 2k + 2R]$, Cramer's Rule implies that each $|a_i|$ is at most $(2k + 2R)(j-1)!(2k)^{j-1}$. It follows that there is a linear combination of the first $j-1$ equations of the final system, which yields $\tau_1 = \sum_{i=j}^{k} \gamma_i^j \tau_i$, where each $|\gamma_i^j|$ is at most $(j-1) \cdot (2k) \cdot (2k + 2R)(j-1)!(2k)^{j-1} \leq (2k + 2R)(2(j-1)k)^j$. Setting $\tau = \tau^*$ in the equation above, we get $\tau_1^* \leq (k - j + 1)(\max_i |\gamma_i^j|)\tau_j^*$ which implies

$$\tau_j^* \geq \frac{\tau_1^*}{(2k + 2R) \cdot (2(j-1)k)^j (k - j + 1)} \geq \frac{\tau_1^*}{(2k + 2R) \cdot (2k)^{2j+1}} \tag{8}$$

Observing that $\sum_{i=1}^{k} \tau_i^* = w_1^*$, we have $\tau_1^* \geq w_1^*/k$, which gives

$$\tau_j^* \geq \frac{w_1^*}{(2k + 2R) \cdot (2k)^{2j+2}}.$$

Finally, we observe that for $j \in [k-1]$, the $j$-th biggest element of the multiset $\Delta_1, \ldots, \Delta_{k-1}$ (see the lemma statement) is at least $\tau_{j+1}$. (It is either $\tau_{j+1}$ or $\tau_j$ depending on whether or not $\delta_k^* = w_k^*$ is among the $j$ largest elements of $\{\delta_1^*, \ldots, \delta_k^*\}$.) Renormalizing so that the largest weight is 1, we have shown that for odd $h'$, the $j$-th biggest element of the multiset $\Delta_1, \ldots, \Delta_{k-1}$ is at least $\frac{1}{(2k+2R) \cdot (2k)^{2j+4}}$. This completes the proof for the case that $h'$ is odd. The extension to general $h'$ is identical to the argument of Lemma 26. This concludes the proof of Lemma 29. □

## 4 Extensions to other distributions

Thus far the notion of approximation that we have dealt with has been approximation under the uniform distribution. In this section we show how our results on small-weight integer approximators can be extended to a fairly broad class of distributions which includes constant-biased product distributions and $K$-wise independent distributions. Our proofs in this more general setting follow the same new approach we have used throughout Section 3, of constructing a "nice" representation and then using anti-concentration.

Formally, given a probability distribution $\mathcal{D}$ on $\{-1,1\}^n$, the *distance between $f, g$ : $\{-1,1\}^n \to \{-1,1\}$ with respect to $\mathcal{D}$* is defined as $\text{dist}_{\mathcal{D}}(f, g) \stackrel{\text{def}}{=} \mathbf{Pr}_{x \leftarrow \mathcal{D}}[f(x) \neq g(x)]$. If $\text{dist}_{\mathcal{D}}(f, g) \leq \epsilon$, we say that $f$ and $g$ are $\epsilon$-close w.r.t. $\mathcal{D}$ and that $g$ is an $\epsilon$-approximator to $f$ (w.r.t. $\mathcal{D}$). We consider the following question: Given a threshold function $f : \{-1,1\}^n \to \{-1,1\}$, an error parameter $\epsilon > 0$ and a distribution $\mathcal{D}$, does there exist an $\epsilon$-approximator $g$ for $f$ w.r.t. $\mathcal{D}$ with small integer weights?

In Section 4.1 we discuss anti-concentration under general distributions and record the anti-concentration inequalities that we will use. In Section 4.2 we generalize the basic $\text{poly}(n) \cdot 2^{\tilde{O}(1/\epsilon^2)}$ result of [Ser07] and in Section 4.3 we generalize the $\text{poly}(n) \cdot 2^{\tilde{O}(1/\epsilon^{2/3})}$ bound of Theorem 2. Because full proofs would be quite lengthy and occasionally repetitive in some cases we only provide sketches.

### 4.1 Anti-concentration under general distributions.

We start by defining the notion of anti-concentration for general measures on the hypercube $\{-1,1\}^n$.

**Definition 31.** *Fix a distribution $\mathcal{D}$ on $\{-1,1\}^n$. Let $a \in \mathbb{R}^n$ be a weight vector and $r \in \mathbb{R}_+$. The Lévy anti-concentration function of $a$ w.r.t. $\mathcal{D}$ is defined as*

$$p_r(a, \mathcal{D}) \stackrel{\text{def}}{=} \sup_{v \in \mathbb{R}} \mathbf{Pr}_{x \leftarrow \mathcal{D}}[|a \cdot x - v| \leq r].$$

Let $x \in \{-1,1\}^n$ be drawn from $\mathcal{D}$ and consider the random variable $S = a \cdot x = \sum_{i=1}^n a_i x_i$ where $a \in \mathbb{R}^n$. While it is clear that the random variable $S$ can be very concentrated if $\mathcal{D}$ is arbitrary, there are broad classes of interesting distributions for which it is possible to establish good anti-concentration under suitable assumptions for the weights. In particular, as we now describe, this is possible for constant-biased product distributions and $K$-wise independent distributions for large enough $K$.

**Product Distributions.** We start with the case of product distributions. Let $p_i \in (0, 1)$, $i \in [n]$. Let $\mu_{p_i}$ be the distribution on the two point space $\{-1, 1\}$ with $\mu_{p_i}(1) = p_i$. We denote the corresponding product distribution by $\bigotimes_{i=1}^n \mu_{p_i}$. For such a product distribution we denote
$$p \stackrel{\text{def}}{=} \min_{j \in [n]}\{p_j, 1 - p_j\} \in (0, 1/2].$$

We henceforth write $\mathcal{D}_{\text{prod}}$ to denote a generic product distribution for which $p = \Theta(1)$ (we call such distributions *constant-biased product distributions*), and we omit the dependence on $p$ in our bounds.

We mention two anti-concentration inequalities under product distributions. The following results intuitively say that, for any constant-bounded product distribution, the random variable $S = a \cdot x = \sum_{i=1}^n a_i x_i$ has good anti-concentration if the weights have appropriate structure.

The first such result is a generalization of Erdős's theorem, Theorem 19:

**Theorem 32.** *Let $a = (a_1, \ldots, a_k) \in \mathbb{R}^k$, $r \in \mathbb{R}_+$ be such that $|a_i| \geq r$ for all $i \in [k]$. Then $p_r(a, \mathcal{D}_{\text{prod}}) \leq O(k^{-1/2})$.*

For the case that $p_i = p$ for all $i \in [k]$, Theorem 32 can be proved in an elementary way using Sperner theory (similar to the proof of Theorem 19). For the case of different $p_i$'s, a proof can be obtained using the Fourier analytic methods in [Hal77] (see [AGKW09] for an explicit reference).

The second theorem is a generalization of Theorem 20:

**Theorem 33.** *Let $a = (a_1, \ldots, a_k) \in \mathbb{R}^k$, $r \in \mathbb{R}_+$ be such that $|a_i - a_j| \geq r$ for all $i \neq j \in [k]$. Then $p_r(a, \mathcal{D}_{\text{prod}}) \leq O(k^{-3/2})$.*

This theorem can also be obtained using the techniques of [Hal77].

Finally, the extension lemma will again be useful for us:

**Lemma 34** (Extension). *Let $a \in \mathbb{R}^k$ be any $k$-dimensional weight vector and $r \in \mathbb{R}_+$ any non-negative real. For any $n > k$, let $a' \in \mathbb{R}^n$ be the vector $(a_1, \ldots, a_k, a'_{k+1}, \ldots, a'_n)$ where the weights $a'_{k+1}, \ldots, a'_n$ may be any real numbers. Then we have $p_r(a', \mathcal{D}_{\text{prod}}) \leq p_r(a, \mathcal{D}_{\text{prod}})$.*

As in the uniform distribution case, this follows directly from independence.

**K-wise Independent Distributions.** A distribution $\mathcal{D}$ on $\{-1, 1\}^n$ is *K-wise independent* if the projection of $\mathcal{D}$ onto any $K$ indices is uniformly distributed over $\{-1, 1\}^K$. The class of $K$-wise independent distributions over $\{-1, 1\}^n$ is a broad and important class of distributions that has received much study (see [Wig94, BR94] and many other references) because of its usefulness in derandomization and other applications.

We note that the extension lemma fails for $K$-wise independent distributions, since one cannot fix most of the bits and argue that the remaining bits are independent. For $K$-wise independent distributions, we thus need to establish anti-concentration in a different way. This can be done using the recent result of [DGJ+09].

We shall denote by $\mathcal{D}_{\text{Kwise}}$ a generic $K$-wise independent distribution on $\{-1, 1\}^n$. We recall the main result of [DGJ+09]:

**Theorem 35** (Theorem 1.2 in [DGJ+09], rephrased)**.** *Let $h(x) = \text{sign}(\sum_{i=1}^{n} w_i x_i - \theta)$ be any threshold function. Then we have*

$$\left| \mathbf{Pr}_{x \leftarrow \mathcal{D}_{\text{Kwise}}}[h(x) = 1] - \mathbf{Pr}_{x \leftarrow \mathcal{U}}[h(x) = 1] \right| \leq O\left(\frac{\log K}{\sqrt{K}}\right).$$

As an immediate corollary we obtain:

**Fact 36.** *Let $a \in \mathbb{R}^n$ and $r \in \mathbb{R}_+$. Then*

$$p_r(a, \mathcal{D}_{\text{Kwise}}) \leq p_r(a, \mathcal{U}) + O\left(\frac{\log K}{\sqrt{K}}\right).$$

We require the above two anti-concentration probabilities to be $\epsilon$-close to each other, so we henceforth fix

$$K \stackrel{\text{def}}{=} \Theta(1/\epsilon^2 \cdot \log^2(1/\epsilon)) = \tilde{O}(1/\epsilon^2).$$

Fact 36, Theorem 19 and Lemma 21 now yield:

**Theorem 37.** *Let $a = (a_1, \ldots, a_n) \in \mathbb{R}^n$, $r \in \mathbb{R}_+$ and suppose that $|a_i| \geq r$ for all $i \in [\ell]$. Then*

$$p_r(a, \mathcal{D}_{\text{Kwise}}) \leq O(\ell^{-1/2}) + \epsilon.$$

Similarly Fact 36, Theorem 20 and Lemma 21 together yield:

**Theorem 38.** *Let $a = (a_1, \ldots, a_n) \in \mathbb{R}^n$, $r \in \mathbb{R}_+$ and suppose that $|a_i - a_j| \geq r$ for all $i \neq j \in [\ell]$. Then*

$$p_r(a, \mathcal{D}_{\text{Kwise}}) \leq O(\ell^{-3/2}) + \epsilon.$$

## 4.2 A $\text{poly}(n) \cdot 2^{\tilde{O}(1/\epsilon^2)}$ bound for product distributions and $K$-wise independent distributions.

The following lemma translates anti-concentration to low-integer weight approximation for *any* distribution $\mathcal{D}$.

**Lemma 39.** *Fix a distribution $\mathcal{D}$ on $\{-1, 1\}^n$. Let $g = \text{sign}(w \cdot x - \theta)$ be any threshold function. If $p_r(w, \mathcal{D}) \leq \epsilon$, then there exists an $\epsilon$-approximator $h$ for $g$ w.r.t. $\mathcal{D}$, where $h$ is a threshold function with integer weights each of magnitude $O(\max_i |w_i| \cdot n/r)$.*

We note that the bound on the magnitude of the weights is now linear in $n$ (as opposed to $\sqrt{n}$ in Lemma 22) and that no dependence on $\epsilon$ appears in the bound. The proof is essentially the same as the proof of Lemma 22 but with the following small change: in Lemma 22, we rounded to integer multiples of $r/\sqrt{n \log(1/\epsilon)}$ and used a Hoeffding bound to show that the probability that the error vector $e$ has $|e \cdot x| \geq r$, for a uniformly random $x \in \{-1, 1\}^n$, is upper bounded by $\epsilon$. Since the Hoeffding bound does not apply for general distributions, we now round to multiples of $r/n$. (This is what makes the dependence on $n$ worse by a $\sqrt{n}$ factor.) As a consequence, the corresponding error probability is now 0, since $\|e\|_1 < r$.

To obtain the desired $\text{poly}(n) \cdot 2^{\tilde{O}(1/\epsilon^2)}$ bound, we will use the following claim, that follows from Claim 23 by setting $\epsilon = 1/(2^n + 1)$. (The claim also follows as an immediate corollary of a theorem, due to Håstad [Hås05], that appears as Theorem 6.5 in [Ser07].)

**Claim 40.** *Let $f : \{-1, 1\}^n \to \{-1, 1\}$ be any threshold function and let $k \in [n]$. There is an exact representation for $f$ as $\text{sign}(\sum_{i=1}^{n} w_i x_i - \theta)$ with the following property: Suppose (reordering and rescaling weights if necessary) that $1 = |w_1| \geq \cdots \geq |w_n|$. Then $|w_k| \geq 1/(4 \cdot k^k \cdot n)$.*

At this point we have the tools to obtain a weight bound of $n^2 \cdot 2^{\tilde{O}(1/\epsilon^2)}$ for both product distributions and $K$-wise independent distributions. However, the following easily verified remarks can be used to improve the dependence on $n$ to linear in both cases.

1. For the class of product distributions, Lemma 39 applies with the same quantitative bound as Lemma 22 (i.e. $O(\max_i |w_i| \cdot \sqrt{n \ln(1/\epsilon)}/r)$). The proof is essentially the same as that of Lemma 22, since the Hoeffding bound only requires independence.

2. For the class of $K$-wise independent distributions, we can obtain the quantitative bound $\max_i |w_i| \cdot (\sqrt{n}/r) \cdot \tilde{O}(1/\epsilon^2)$ in Lemma 39. The proof is essentially the same as Lemma 22 but using an appropriate tail inequality for $K$-wise independent distributions (e.g. Lemma 2.3 in [BR94]) instead of the Hoeffding bound.

3. Claim 23 applies with the same quantitative bound for any constant-biased product distribution and with the bound $1/(k^k \cdot \sqrt{n} \cdot \tilde{O}(1/\epsilon^2))$ for any $K$-wise independent distribution. For product distributions the proof remains essentially unchanged from the uniform distribution proof in [OS08], again because the Hoeffding bound only requires independence. For $K$-wise independent distributions, again one uses a tail bound for $K$-wise independent distributions.

Thus we have the main results of this subsection:

**Theorem 41.** *Let $f$ be any $n$-variable threshold function. Then*

1. *$f$ is $\epsilon$-approximated with respect to $\mathcal{D}_{\text{prod}}$ by a threshold function $g = \text{sign}(w \cdot x - \theta)$ with $w_1, \ldots, w_n$ all integers of magnitude $n \cdot 2^{\tilde{O}(1/\epsilon^2)}$; and*

2. *$f$ is $\epsilon$-approximated with respect to $\mathcal{D}_{\text{Kwise}}$ by a threshold function $g = \text{sign}(w \cdot x - \theta)$ with $w_1, \ldots, w_n$ all integers of magnitude $n \cdot 2^{\tilde{O}(1/\epsilon^2)}$.*

*Proof.* For part (1), we set $\ell = \min\{1/\epsilon^2, n\}$. We apply Theorem 32 to the weight vector $(w_1, \ldots, w_\ell)$ from Claim 23 (or more precisely from the variant described in remark 3 above) taking $r = (1/\sqrt{n}) \cdot 2^{-\tilde{O}(1/\epsilon^2)}$. Theorem 32 gives $p_r((w_1, \ldots, w_l), \mathcal{D}_{\text{prod}}) \leq \epsilon$, and Lemma 34 gives $p_r(w, \mathcal{D}_{\text{prod}}) \leq \epsilon$. An application of Lemma 39 (see remark 1 above) completes the proof.

For part (2), we set $\ell = \min\{1/\epsilon^2, n\}$. We apply Theorem 37 to the weight vector $(w_1, \ldots, w_\ell)$ from the modified Claim 23 (see remark 3, above), taking $r = (1/\sqrt{n}) \cdot 2^{\tilde{O}(1/\epsilon^2)}$. Theorem 37 directly yields $p_r(w, \mathcal{D}_{\text{Kwise}}) \leq 2\epsilon$. An application of Lemma 39 (see remark 2 above) completes the proof. □

## 4.3 A $\text{poly}(n) \cdot 2^{\tilde{O}(1/\epsilon^{2/3})}$ bound for product distributions and $K$-wise independent distributions.

In this subsection we first prove a bound of $n^{O(1/\epsilon^{2/3})}$ (analogous to Section 3.3). We then sketch how the arguments of [Ser07] can be extended to the product distribution and $K$-wise independent distribution settings to obtain the final result (analogous to Section 3.4).

### 4.3.1 An $n^{O(1/\epsilon^{2/3})}$ bound.

With the required machinery in place it is straightforward to prove:

**Theorem 42.** *Let $f$ be any $n$-variable threshold function. Then:*

1. *$f$ is $\epsilon$-approximated with respect to $\mathcal{D}_{\text{prod}}$ by a threshold function $g = \text{sign}(w \cdot x - \theta)$ with $w_1, \ldots, w_n$ all integers of magnitude $n^{O(1/\epsilon^{2/3})}$.*

2. *$f$ is $\epsilon$-approximated with respect to $\mathcal{D}_{\text{Kwise}}$ by a threshold function $g = \text{sign}(w \cdot x - \theta)$ with $w_1, \ldots, w_n$ all integers of magnitude $n^{O(1/\epsilon^{2/3})}$.*

*Proof.* For (1), we set $\ell = \min\{1/\epsilon^{2/3}, n\}$. We apply Theorem 33 to the weight vector $(w_{i_1}, \ldots, w_{i_\ell})$ obtained from Lemma 26, taking $r = 1/(2n+2)^{2\ell+8} = n^{-\Omega(1/\epsilon^{2/3})}$. Theorem 33 gives $p_r((w_{i_1}, \ldots, w_{i_\ell}), \mathcal{D}_{\text{prod}}) \leq O(\epsilon)$, and Lemma 34 gives $p_r(w, \mathcal{D}_{\text{prod}}) \leq O(\epsilon)$. An application of Lemma 39 completes the proof.

For (2), we again set $\ell = \min\{1/\epsilon^{2/3}, n\}$. We apply Theorem 38 to the weight vector $(w_{i_1}, \ldots, w_{i_\ell})$ from Lemma 26, taking $r = 1/(2n+2)^{2\ell+8} = n^{-\Omega(1/\epsilon^{2/3})}$. Theorem 37 directly yields $p_r(w, \mathcal{D}_{\text{Kwise}}) \leq 2\epsilon$. An application of Lemma 39 completes the proof. □

### 4.3.2 Completing the proof: a $\text{poly}(n) \cdot 2^{\tilde{O}(1/\epsilon^{2/3})}$ bound.

Finally, in this section we show how to obtain a $\text{poly}(n) \cdot 2^{\tilde{O}(1/\epsilon^{2/3})}$ bound for constant-biased product distributions and for $K$-wise independent distributions:

**Theorem 43.** *Let $\mathcal{D}$ be either a constant-biased product distribution or a $K$-wise independent distribution. Every $n$-variable threshold function $f$ is $\epsilon$-approximated w.r.t. $\mathcal{D}$ by a threshold function $g = \text{sign}(w \cdot x - \theta)$ with $w_1, \ldots, w_n$ all integers of magnitude $\text{poly}(n) \cdot 2^{\tilde{O}(1/\epsilon^{2/3})}$.*

The theorem is proved following an approach analogous to Section 3.4. To do this, one needs to check that the ideas from [Ser07] can be appropriately generalized to product distributions and $K$-wise independent distributions. We do not present the details of the proofs but only briefly sketch the ingredients that make these generalizations possible.

Fix a distribution $\mathcal{D}$ in either of the aforementioned classes. Let $f$ be a threshold function and $\epsilon > 0$ be given. The first thing one must argue is that for an appropriate threshold $L = \tilde{O}(1/\epsilon^2)$, if the $\epsilon$-critical index $\ell$ is bigger than $L$, then $f$ is $\epsilon$-close *with respect to $\mathcal{D}$* to the $L$-junta $g$ obtained by truncating the smallest $n - L$ weights. If $\mathcal{D}$ is a constant-biased product distribution, this can be done by an analysis very similar to Case IIa in [Ser07]. The only difference is in constant factors that eventually lead the threshold $L$ to increase by a factor of $1/p$ (the bias of the distribution). If $\mathcal{D}$ is a $K$-wise independent distribution, we can no longer use the Hoeffding bound that is used in Case IIa from [Ser07]. However, it turns out that Chebyshev's bound can be used instead; indeed, [DGJ$^+$09] does precisely this to show that if the $\epsilon$-critical index is greater than $L$, then $f$ is $\epsilon$-close to the $L$-junta $g$. w.r.t. any $K$-wise independent distribution.

To conclude the sketch it suffices to argue that, for every value of $\ell$, there exists an $\epsilon$-approximator for $f$ (w.r.t. $\mathcal{D}$) with integer weights of magnitude $\text{poly}(n/\epsilon) \cdot 2^{O(\ell \log \ell)}$. A generalization of Case IIb from [Ser07] is also possible in this case. For constant-biased product distributions this is easy (in fact we obtain a bound of $\sqrt{n \log(1/\epsilon)} 2^{O(\ell \log \ell)}$), because the Hoeffding bound and Gaussian anti-concentration still apply. For $K$-wise distributions, again one can use the tail bounds in [BR94] for $K$-wise independent distributions in place of the Hoeffding bound; Gaussian anti-concentration can also be deduced in this case using Fact 36.

Now using Lemma 29, the proof follows as in Section 3.4.

## 4.4 Discussion: some distributions require large-weight approximators.

We have shown that for some non-uniform distributions such as constant-biased product distributions and $K$-wise independent distributions, every threshold function can be $\epsilon$-approximated using integer weights at most $\text{poly}(n) \cdot 2^{\tilde{O}(1/\epsilon^{2/3})}$. An optimist might wonder whether it is possible that under *any* distribution $\mathcal{D}$, every threshold function can be $\epsilon$-approximated with integer weights $\text{poly}(n) \cdot 2^{\tilde{O}(1/\epsilon^{2/3})}$, or perhaps $n^{\tilde{O}(1/\epsilon^{2/3})}$. Here we observe that such a strong bound cannot hold for every distribution:

**Proposition 44.** *There is a probability distribution $\mathcal{D}$ over $\{-1, 1\}^n$ and a threshold function $f$ such that any integer-weight threshold function that $1/(n+2)$-approximates $f$ under $\mathcal{D}$ must have weight $2^{\Omega(n)}$.*

*Proof.* The function $f$ is the "ODD-MAX-BIT" function [Bei94] which, on input $x$, outputs $(-1)^i$ where $i$ is the first index such that $x_i = 1$. It is straightforward to verify that $f$ is a threshold function, and it is well known that any integer-weight representation of $f$ must have weight $2^{\Omega(n)}$ (see e.g. [HV86]).

Anthony et al. [ABST95] give an explicit set $S$ of $n+1$ points from $\{-1, 1\}^n$ and show that any threshold function $h$ that agrees with $f$ on all $n+1$ points in $S$ must in fact be identical to $f$ on all of $\{-1, 1\}^n$ (the set $S$ is said to be a "specifying set" for $f$). Under the uniform distribution on $S$ any $1/(n+2)$-approximator must be correct on all points of $S$, and hence identical to $f$, and the result follows. □

# 5 Conclusions and Future Work

We have already discussed directions for future work relating to Theorem 1 in Section 2.5. Regarding Theorem 2, we feel that our high-level approach using anti-concentration holds promise for substantial further progress. Significant strengthenings of Halász's anti-concentration bound are known under stronger restrictions on the additive structure of the weights $w_1, \ldots, w_n$, see e.g. [Vu08, TV08]. Can corresponding extensions of Lemma 26 be established, proving that every threshold function admits a representation with weights that have the required structure? Perhaps every threshold function $f$ can be $\epsilon$-approximated using integer weights at most $\text{poly}(n) \cdot 2^{\text{polylog}(1/\epsilon)}$. We hope that further study of our anti-concentration based approach may yield such a bound.

**Acknowledgements.** We thank Ryan O'Donnell for asking a question that led to Theorem 1.

[Vu08]      V. Vu. A structural approach to subset-sum problems. Available at http://arxiv.org/abs/0804.3211, 2008.

[Wig94]      A. Wigderson. The amazing power of pairwise independence. In *Proceedings of the 26th ACM Symposium on Theory of Computing*, pages 645–647, 1994.

# Appendix

### Proof of Claim 27.

Recall Claim 27:

**Claim 27.** *There exists a set of $n$ points $y^{(1)}, \ldots, y^{(n)} \in f^{-1}(1) \cup D_n$ such that $w^*$ is the unique solution of the linear system $\{w \cdot y^{(i)} = 1 \mid i = 1, 2, \ldots, n\}$, denoted (\*).*

*Proof.* By definition, $w^* \in \mathbb{R}^n$ is a weight vector that satisfies a *maximum* number of constraints in $\mathcal{LP}$ with equality. That is, there exist $s \in \mathbb{N}^*$ input vectors $y^{(i_1)}, y^{(i_2)}, \ldots, y^{(i_s)} \in f^{-1}(1) \cup D_n$ such that

$$\begin{pmatrix} y^{(i_1)} \\ y^{(i_2)} \\ \ldots \\ y^{(i_s)} \end{pmatrix} w^* = A w^* = \mathbf{1}_{n \times 1} \tag{9}$$

and no other weight vector satisfies more than $s$ of the constraints with equality. Also note that for all input vectors $x \in f^{-1}(1) \cup D_n \setminus \{y^{(i_j)}\}_{j \in [s]}$ it holds $w^* \cdot x > 1$.

Consider the linear system $Aw = \mathbf{1}_{n \times 1}$. By definition, $w^*$ is a solution to this system. We will show that the system has a unique solution or equivalently that $\mathrm{rank}(A) = n$. Then by selecting $n$ linearly independent rows of the matrix $A$, we get the linear system (\*).

Suppose, for the sake of contradiction, that $\mathrm{rank}(A) < n$. Then, there exists a non-zero vector that lies in the right null-space of $A$, i.e. there exists $\mathbf{0}_{n \times 1} \neq u \in \mathbb{R}^n$ such that $Au = \mathbf{0}_{s \times 1}$. Consider the family of weight vectors $\{w_\epsilon^* \stackrel{\mathrm{def}}{=} w^* + \epsilon u\}_{\epsilon \in \mathbb{R}}$. We will argue that there exists $\epsilon_0 \in \mathbb{R}^*$ such that the vector $w_{\epsilon_0}^* \neq w^*$ satisfies at least $s+1$ of the constraints of $\mathcal{LP}$ with equality, which is a contradiction.

We now proceed with the argument. We have the following:

1. For all $\epsilon \in \mathbb{R}$ and for all $j \in [s]$ it holds $w_\epsilon^* \cdot y^{(i_j)} = w^* \cdot y^{(i_j)} + \epsilon(u \cdot y^{(i_j)}) = 1$, since $u \cdot y^{(i_j)} = 0$, by (9).

2. There exists at least one vector $y \in (f^{-1}(1) \cup D_n) \setminus \{y^{(i_j)}\}_{j \in [s]}$ such that $u \cdot y \neq 0$. This holds true because the set $f^{-1}(1) \cup D_n$ (in fact $f^{-1}(1)$ itself) spans $\mathbb{R}^n$, while we are assuming that the rank of $A$ is strictly less than $n$. (Recall that $f$ was assumed to be odd, hence $f^{-1}(1)$ contains either $x$ or $-x$ for every $x \in \{-1, 1\}^n$.) Let $U \neq \emptyset$ be the corresponding set, i.e. $U \stackrel{\mathrm{def}}{=} \{y \in f^{-1}(1) \cup D_n \setminus \{y^{(i_j)}\}_{j \in [s]} \mid u \cdot y \neq 0\}$. Let us also denote $\overline{U} \stackrel{\mathrm{def}}{=} \{y \in f^{-1}(1) \cup D_n \setminus \{y^{(i_j)}\}_{j \in [s]} \mid u \cdot y = 0\}$ for its complement.

We now claim that one can choose an appropriate value $\epsilon_0$ for $\epsilon$ such that for some $y' \in f^{-1}(1) \cup D_n \setminus \{y^{(i_j)}\}_{j \in [s]}$ we have $w^*_{\epsilon_0} \cdot y' = 1$ and for all $x \in f^{-1}(1) \cup D_n \setminus \{y^{(i_j)}\}_{j \in [s]}$ it holds $w^*_{\epsilon_0} \cdot x \geq 1$. The latter statement provides the desired contradiction, since, combined with (1) above, it implies that the corresponding vector $w^*_{\epsilon_0}$ is a feasible solution to $\mathcal{LP}$ and satisfies (at least) $s+1$ constraints with equality – in particular, those corresponding to the points $\{y^{(i_j)}\}_{j \in [s]} \cup \{y'\}$.

Partition the set $U$ into $U_+ \overset{\text{def}}{=} \{y \in f^{-1}(1) \cup D_n \setminus \{y^{(i_j)}\}_{j \in [s]} \mid u \cdot y > 0\}$ and $U_- \overset{\text{def}}{=} U \setminus U_+$. By (2) above, we know that at least one of the sets $U_+$ and $U_-$ is nonempty. We analyze the case that $U_+ \neq \emptyset$, the case $U_- \neq \emptyset$ being very similar. Recall that for every $x \in f^{-1}(1) \cup D_n \setminus \{y^{(i_j)}\}_{j \in [s]}$ it holds $w^* \cdot x > 1$. Now consider some $x_+ \in U_+$; we have that $w^* \cdot x_+ > 1$ and $u \cdot x_+ > 0$. We therefore select:

$$\epsilon_0 \overset{\text{def}}{=} \max_{x_+ \in U_+} \frac{1 - w^* \cdot x_+}{u \cdot x_+} \tag{10}$$

First, it is clear that $\epsilon_0 < 0$, which implies that $w^*_{\epsilon_0} \neq w^*$. It is also straightforward to verify that the remaining desired properties are satisfied. Indeed, there exists at least one point $y' \in U_+ \subseteq f^{-1}(1) \cup D_n \setminus \{y^{(i_j)}\}_{j \in [s]}$ – a maximizer of (10) – such that $w^*_{\epsilon_0} \cdot y' = w^* \cdot y' + \epsilon_0 (u \cdot y') = 1$. Also, if $x \in U_+$, then by the definition of $\epsilon_0$ above, we have that $1 \leq w^*_{\epsilon_0} \cdot x < w^* \cdot x$. Now if $x \in U_-$, then $w^*_{\epsilon_0} \cdot x > w^* \cdot x > 1$. Finally, if $x \in \overline{U}$, then $w^*_{\epsilon_0} \cdot x = w^* \cdot x > 1$. Hence, we have $w^*_{\epsilon_0} \cdot x \geq 1$ for all $x \in f^{-1}(1) \cup D_n \setminus \{y^{(i_j)}\}_{j \in [s]}$ which completes the proof of Claim 27. $\square$